\documentclass[aps,prl,twocolumn,superscriptaddress,showpacs,floatfix,nofootinbib]{revtex4-1}
\usepackage{amsmath,graphicx}
\usepackage{hyperref}
\usepackage{orcidlink}
\usepackage[x11names]{xcolor}
\usepackage{bm}
\usepackage{amsfonts,amssymb}

\def\TRENTo{{\sc t\kern-.05em \lower.5ex\hbox{r}\kern-.025em e\kern-.05em n\kern-.05em t\kern-.09em}o}
\def\iccing{{\sc i\kern-.05em c\kern-.05em c\kern-.05em i\kern-.05em n\kern-.05em g\kern-.05em}}
\def\ccake{{\sc c\kern-.05em c\kern-.05em a\kern-.05em k\kern-.05em e\kern-.05em}}
\def\vUSPhydro{v-{\sc u\kern-.05em s\kern-.05em p\kern-.05em}hydro}
\graphicspath{ {figures/} }

\begin{document}

\title{Symmetry-Energy Expansion with Strange Dense Matter}

\author{Yumu 
Yang\,\orcidlink{0009-0001-8979-9343}}
\email{yumuy2@illinois.edu}
\affiliation{The Grainger College of Engineering, Illinois Center for Advanced Studies of the Universe, Department of Physics, University of Illinois at Urbana-Champaign, Urbana, IL 61801, USA}
\author{Nikolas Cruz Camacho\,\orcidlink{0009-0004-7870-0039}}
\email{cnc6@illinois.edu}
\affiliation{The Grainger College of Engineering, Illinois Center for Advanced Studies of the Universe, Department of Physics, University of Illinois at Urbana-Champaign, Urbana, IL 61801, USA}
\author{Mauricio Hippert\,\orcidlink{0000-0001-5802-3908}}
\email{hippert@cbpf.br}
\affiliation{Centro Brasileiro de Pesquisas Físicas, Rua Dr. Xavier Sigaud 150, 
Rio de Janeiro, RJ, 22290-180, Brazil}
\author{Jacquelyn Noronha-Hostler\,\orcidlink{0000-0003-3229-4958}}
\email{jnorhos@illinois.edu}
\affiliation{The Grainger College of Engineering, Illinois Center for Advanced Studies of the Universe, Department of Physics, University of Illinois at Urbana-Champaign, Urbana, IL 61801, USA}

\date{\today}

\begin{abstract}
The quantum chromodynamics (QCD) phase diagram at large densities and low temperatures can be probed using both neutron stars and low-energy heavy-ion collisions. Heavy-ion collisions are nearly isospin-symmetric systems, whereas neutron stars are highly isospin asymmetric since they are neutron rich. The symmetry-energy expansion is used to connect these regimes across isospin asymmetry. However, the current symmetry-energy expansion does not account for strange particles. 
In this work, we include finite strangeness by redefining the isospin-asymmetry parameter and the symmetry-energy expansion in a way that is consistent with QCD SU(3) flavor symmetry. Our new symmetry energy works well beyond typical neutron star central densities and admits a skewness term in the presence of strangeness for the case of weak equilibrium.
\end{abstract}

\maketitle

\section{Introduction}
Depending on the temperatures $T$, baryon densities $n_B$, electric charge densities $n_Q$, and strangeness densities $n_S$ reached, nuclear matter may have drastically different features \cite{Collins:1974ky,Cabibbo:1975ig,Bzdak:2019pkr,An:2021wof,Sorensen:2023zkk,Lovato:2022vgq}. 
In the laboratory, one generally discusses the charge fraction $Y_Q=Z/A=n_Q/n_B$ because it is dictated by the choice of nuclei with proton number $Z$ and total number of nucleons, $A$. Similarly, one can define a strangeness fraction of $Y_S=n_S/n_B$.
The quantum chromodynamics (QCD) phase diagram at finite density is a four-dimensional $\left\{T,n_B,Y_Q,Y_S\right\}$ space that can be probed by laboratory experiments (heavy-ion collisions \cite{Danielewicz:2002pu,Pratt:2015zsa,CBM:2016kpk,Alba:2017hhe,Spieles:2020zaa,Monnai:2021kgu,Oliinychenko:2022uvy,Yao:2023yda,Plumberg:2024leb}), astrophysical objects (neutron stars \cite{LIGOScientific:2017ync,Miller:2019cac,Miller:2021qha,Riley:2019yda,Riley:2021pdl}), and astrophysical events (binary neutron star mergers \cite{Most:2018eaw}, supernovae \cite{Zha:2020gjw}, or even the big bang \cite{Borsanyi:2016ksw}).

In the low-$T$ and large-$n_B$ limit, we have three different systems to probe the phase diagram: (i) neutron stars at low $T\approx 0$ with small $Y_Q\lesssim 0.1$, (ii) neutron star mergers at $T\approx [0,100] \, \rm{MeV}$ with $Y_Q \approx [0.01,0.2]$, and (iii) low-energy heavy-ion collisions at $T\approx [50,150] \, \rm{MeV}$ with $Y_Q\approx [0.38,0.5]$. 
Values of $Y_S$ are often excluded from discussions of the phase diagram because heavy-ion collisions probe global $Y_S=0$ (due to strangeness conservation in QCD) \cite{Monnai:2019hkn,Noronha-Hostler:2019ayj,Karthein:2021nxe},%
\footnote{This does not exclude local fluctuations in $Y_S$ \cite{Xu:2016hxf, Noronha-Hostler:2016rpd, Bluhm:2018aei, Alba:2017mqu, Bellwied:2018tkc, Bellwied:2019pxh, Alba:2020jir, SanMartin:2023zhv}.} 
and because, even though an average $Y_S<0$ is allowed by weak interactions in neutron stars, the presence of strangeness in neutron star matter is not settled.%
\footnote{If strangeness is present, then  $Y_S< 0$ is expected, since strange quarks carry negative strangeness and the strange antiquark is suppressed due to its negative baryon number.} 
Confirming the presence of strange particles within the core of neutron stars is of great importance to the nuclear physics community \cite{Kaplan:1986yq, Glendenning:1992vb, Schaffner:1993qj, Schaffner:1993nn, Hatsuda:1994pi, Pisarski:1999gq, Buballa:2003qv, Weber:2004kj, Weissenborn:2011kb, Dexheimer:2014pea, Gomes:2014aka, Tolos:2020aln}, as it would establish the first known regime with a stable $Y_S\neq 0$ in the universe. 
In fact, phase transitions can become significantly more complicated when multiple conserved charges are present \cite{Glendenning:1992vb}.
However, determining the properties at the core of a neutron star is not yet possible with current astrophysical data \cite{Mroczek:2023zxo}. 

Thus, forging phenomenological tools to connect neutron stars to heavy-ion collisions is necessary to provide further constraints to the neutron star equation of state (EOS) to determine its interior structure (see, e.g., Ref.~\cite{Yao:2023yda}). 
Already, the symmetry-energy expansion \cite{Wiringa:1988tp, Steiner:2006bx, Chen:2007ih, Vidana:2009is, Hebeler:2010jx, Cai:2011zn, Drischler:2013iza, Seif:2013tja, Gandolfi:2013baa, Drischler:2015eba, Wellenhofer:2015qba, Wellenhofer:2016lnl, Drischler:2017wtt, Zhang:2018vrx, Li:2019xxz, Wen:2020nqs, Drischler:2020hwi, Somasundaram:2020chb, Imam:2021dbe, Drischler:2021kxf, Sun:2023xkg} can connect the limits of isospin-symmetric nuclear matter (SNM) $Y_Q=0.5$ and pure neutron matter (PNM) $Y_Q=0$ to isospin-asymmetric nuclear matter (ANM) found in neutron stars, where $Y_Q(n_B)$ has a functional form that depends on $n_B$, determined by $\beta$ equilibrium.
The symmetry-energy expansion has been used both to connect estimates for the EOS in heavy-ion collisions and in neutron stars and to extend theoretical results (such as the EOS from chiral effective field theory \cite{Wellenhofer:2015qba,Wellenhofer:2016lnl,Drischler:2020hwi,Drischler:2021kxf}) from the SNM and PNM limits to the case of neutron star matter at $\beta$ equilibrium. 
Only recently \cite{Yao:2023yda}, this expansion has been used to convert a family of neutron star EOS to the SNM regime and to then test them against heavy-ion flow data (while eliminating many EOS through causality, stability, and saturation property constraints), opening up a new path to further constrain the neutron star EOS. 

However, the usual symmetry-energy expansion cannot handle $Y_S\neq 0$, such that it cannot be used with current constraints from heavy ions or neutron stars to test the possibility of strange particles within neutron stars. 
Some groups have attempted to investigate the relation between the strangeness content and the symmetry energy \cite{Providencia:2012rx, Providencia:2013dsa}, or to consider the asymmetry between protons and neutrons in the presence of lambdas without considering isospin-carrying hyperons \cite{Bednarek:2019xytNI}, but did not systematically consider the influence of strangeness on isospin asymmetry. 
That being said, there is a straightforward method to incorporate $Y_S\neq 0$ into the symmetry-energy expansion based on its contribution to isospin, letting us take advantage of the approximate isospin symmetry, which we do for the first time in the work\footnote{Previous work did study this within a hadron resonance gas \cite{Mekjian:2007zz, Mekjian:2007mz} but not on the symmetry-energy expansion}. 
Once strangeness is present, isospin is modified according to the Gell-Mann-Nishijima formula \cite{Nakano:1953zz, Gell-Mann:1956iqa}, which can be directly derived from the SU(3) symmetry of QCD, and we show how the isospin-asymmetry parameter should be correctly defined with strangeness in the next section.

In this work, we consider two specific cases:
an isospin-reflection-symmetric thermodynamic state vs the weak equilibrium case where the strangeness chemical potential vanishes, $\mu_S=0$. 
We then show that our method can reproduce the chiral mean field (CMF) model \cite{Dexheimer:2008ax, Dexheimer:2009hi, Cruz-Camacho:2024odu} results at $T=0$ as one changes the $Y_Q$. 
We discuss new theoretical and experimental methods to put constraints on the expansion coefficients in the strange symmetry-energy expansion.

\textit{Notation.}
We define chemical potentials in the basis of baryon number $B$, charge $Q$, and strangeness $\mathcal{S}$ as%
\footnote{Note that $\mu_B$ and $\mu_S$ become different if isospin is fixed instead of charge. See Ref.~\cite{Aryal:2020ocm} and the Supplemental Material for details~\cite{supplemental}.}
\begin{align}
    & \mu_B\equiv \left(\frac{\partial E}{\partial B}\right)_{Q,\mathcal{S}}, &
    & \mu_Q\equiv\left(\frac{\partial E}{\partial Q}\right)_{B,\mathcal{S}}, &
    & \mu_S\equiv \left(\frac{\partial E}{\partial \mathcal{S}}\right)_{B,Q},  &
    \label{eq:mudefs}
\end{align}
where $E$ is the internal energy of nuclear matter. 
For a given particle $i$, an effective chemical potential can be defined such that
\begin{equation}
\tilde{\mu}_i=B_i\mu_B+S_i\mu_S+Q_i\mu_Q\label{eqn:muQbasis}.
\end{equation}
The isospin along the neutron-proton direction can be related to the conventional $(B,Q,\mathcal{S})$ basis through the Gell-Mann-Nishijima formula \cite{Nakano:1953zz,Gell-Mann:1956iqa}\footnote{In principle, other flavors could be added but we do not expect charm or any heavy quarks to appear within neutron stars.},
\begin{equation}
    Q=I_z+\frac{1}{2}\left(B+\mathcal{S}\right),\label{eqn:GMN}
\end{equation}
which can be used to rewrite Eq.~\eqref{eqn:muQbasis} in the $(B,I_z,\mathcal{S})$ basis \cite{Aryal:2021ojz},
\begin{equation}
\begin{aligned}
    \tilde{\mu}_i &= B_i\left(\mu_B+\frac{1}{2}\mu_Q\right)+S_i\left(\mu_S+\frac{1}{2}\mu_Q\right)+I_{z,i}\mu_Q\label{eqn:muIbasis} \\
    &= B_i\mu_B^{(S,I_z)}+S_i\mu_S^{(B,I_z)}+I_{z,i}\mu_I, 
\end{aligned}
\end{equation}
where 
\begin{subequations}
    \begin{align}
        \mu_B^{(S,I_z)} &\equiv \left(\frac{\partial E}{\partial \mathcal{B}}\right)_{S,I_z} = \mu_B + \frac{1}{2}\mu_Q, \\
        \mu_S^{(B,I_z)} &\equiv \left(\frac{\partial E}{\partial \mathcal{S}}\right)_{B,I_z} = \mu_S + \frac{1}{2}\mu_Q, \\
        \mu_I &\equiv \left(\frac{\partial E}{\partial I_z}\right)_{B,\mathcal{S}} = \mu_Q, 
    \end{align}
\end{subequations}
such that one can choose to write the chemical potentials either in the $BSQ$ basis of $\left\{\mu_B,\mu_S,\mu_Q\right\}$ or the $BSI_z$ basis of $\left\{\mu_B^{(S,I_z)},\mu_S^{(B,I_z)},\mu_I\right\}$. 

\medskip
\section{Isospin Asymmetry with $Y_S\neq 0$}
In a system with more than one conserved charge, charge conjugation symmetry requires that all charges be reflected. 
In the context of QCD, this yields $E(B,Q,\mathcal{S})=E(-B,-Q,-\mathcal{S})$, 
which is not very helpful when we are interested in large positive values of the baryon number only.%
\footnote{This symmetry becomes relevant for lower values of the baryon density, such as can be found in heavy-ion collisions.} 
However, thanks to the approximate isospin symmetry of QCD, another relationship can be found:
$E(B,I_z,\mathcal{S}) \approx E(B,-I_z,\mathcal{S})$, which holds to very good approximation.  
This approximate isospin-flipping $\mathbb{Z}_2$ symmetry of $E$ is more relevant for large values of $B$ and simplifies the symmetry-energy expansion. 

Similar to how Eq.~\eqref{eqn:GMN} relates the charges, one can divide Eq.~\eqref{eqn:GMN} by $B$ to obtain a relation between charge fractions \cite{Aryal:2021ojz}
\begin{equation}
    Y_I=Y_Q-\frac{1}{2}-\frac{1}{2}Y_{\mathcal{S}}.
\end{equation}
In the absence of strangeness, $Y_S=0$, isospin reversal symmetry $I_z \to -I_z$ is equivalent to
\begin{align}\label{eqn:deltaQ}
   & \delta_Q \to -\delta_Q, &
   & \delta_Q \equiv 1-2\,Y_Q, &
   & (Y_S = 0), & 
\end{align}
which is the traditional isospin-asymmetry parameter. 
This reflection symmetry is related to the absence of odd terms in $\delta_Q$ in the usual symmetry-energy expansion, but fails for $Y_S\neq 0$. 
Thus, in the presence of strangeness, we have instead the reflection symmetry
\begin{align}\label{eqn:deltaI}
  &\delta_I\to - \delta_I,&  
  &\delta_I \equiv 1 + Y_S - 2\,Y_Q = -2\,Y_I,&
  &(Y_S\neq 0)&
\end{align}
which generalizes Eq.~\eqref{eqn:deltaQ}.

The inclusion of $Y_S$ in Eq.~\eqref{eqn:deltaI} is necessary for two reasons: 
i) the isospin reversal $I_z \to -I_z$ operation does not commute with the change of basis $(B,I_z,\mathcal{S}) \to (B,Q,\mathcal{S})$ (see the Supplemental Material~\cite{supplemental}), 
and ii) unlike isospin, there is no (approximate) reversal symmetry $\mathcal{S} \to -\mathcal{S}$, unless one also takes $B\to -B$, such that the reflection symmetry under $\delta\to-\delta$ is lost if $Y_S$ is not included in the definition of $\delta$. 
The latter issue is a consequence of the fact that baryons carry negative strangeness. 
Ultimately, this is due to the transformation properties of the representations of $SU(3)$ under charge conjugation.

\begin{figure}
    \centering
    \includegraphics[width=\linewidth]{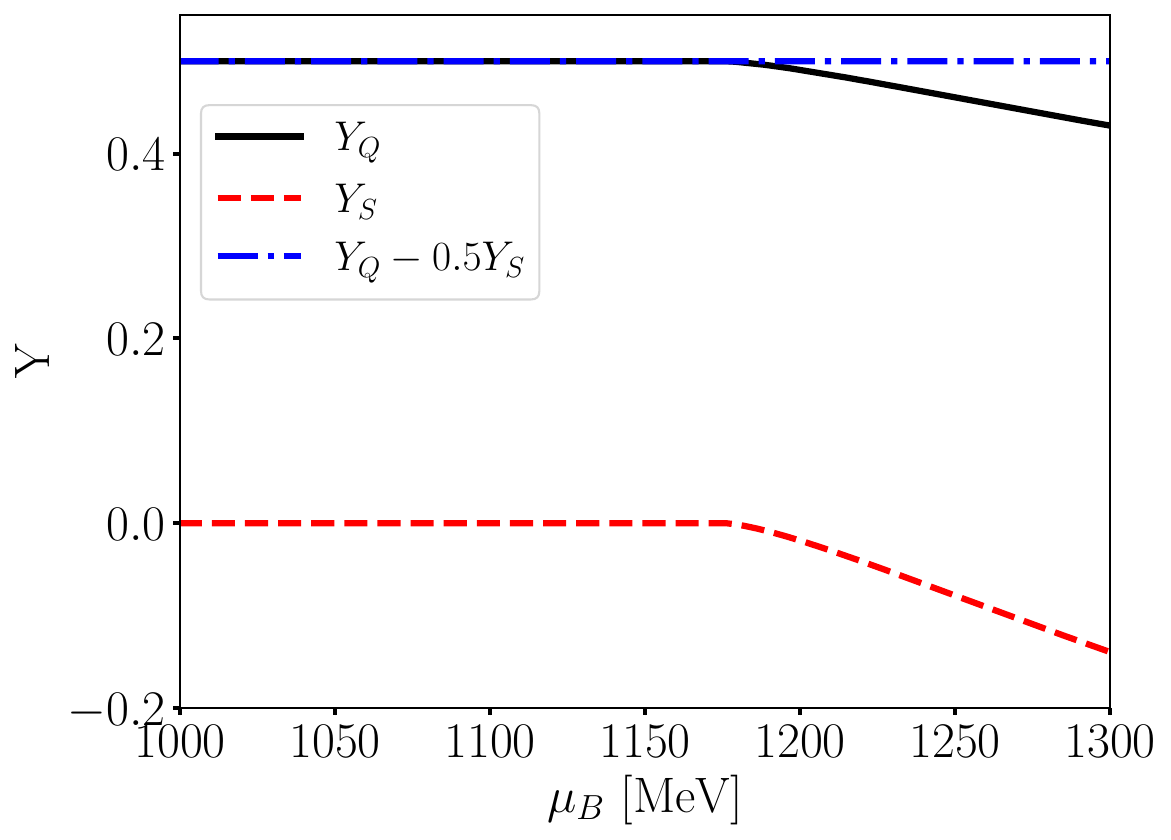}
    \caption{Charge fractions of SNM for $T=\mu_Q=\mu_S=0$. The solid black line is the charge fraction $Y_Q(\mu_B)$, the dashed red line is the strangeness fraction $Y_S(\mu_B)$, and the dot-dashed blue line combines the charge fraction and strangeness contributions to properly account for isospin symmetry. }
    \label{fig:Y_of_SNM}
\end{figure}

The parameter $\delta_I$ solves other issues with the usual $\delta_Q$ parameter in the case where protons are not favored over neutrons, i.e., for vanishing isospin chemical potential. 
For instance, for matter composed of neutrons and protons only,  $Y_Q=\frac{n_p}{n_p+n_n} =\frac{1}{2}$ for  $\mu_I=0$, yielding zero asymmetry, $\delta_Q=0$, as expected. 
However, this fails in the presence of other (strange) particles. 
For instance, in the presence of $\Lambda$  hyperons, 
$ Y_Q=\frac{n_p}{n_p+n_n+n_\Lambda} < \frac{1}{2}$, 
such that  $\delta_Q > 0$, even for $\mu_I=0$. 
If $\delta_I$ is used instead, this problem is solved and $\delta_I=0$ for $\mu_I=0$, even in the presence of neutral and charged hyperons.

Figure\ \ref{fig:Y_of_SNM} illustrates the behavior of  $Y_Q$ as a function of the baryon chemical potential $\mu_B$ for vanishing temperature and $\mu_Q=T=0$. 
The displayed results were obtained for  $\mu_S=0$ (in weak equilibrium) with the recent CMF++ code \cite{Cruz-Camacho:2024odu} for C2 coupling used within the MUSES Collaboration \cite{ReinkePelicer:2025vuh}, which includes the full SU(3) baryon octet and decuplet as well as quarks. 
We find that the change in $Y_Q<1/2$ occurs precisely when $Y_S\neq 0$. 
If we instead consider the combination  $Y_Q-Y_S/2$ at $\mu_Q=0$, then this change is absent and its value remains constant at $1/2$, as illustrated by the blue dot-dashed curve in Fig.\ \ref{fig:Y_of_SNM}. 

 \begin{figure}
    \centering
    \includegraphics[width=\linewidth]{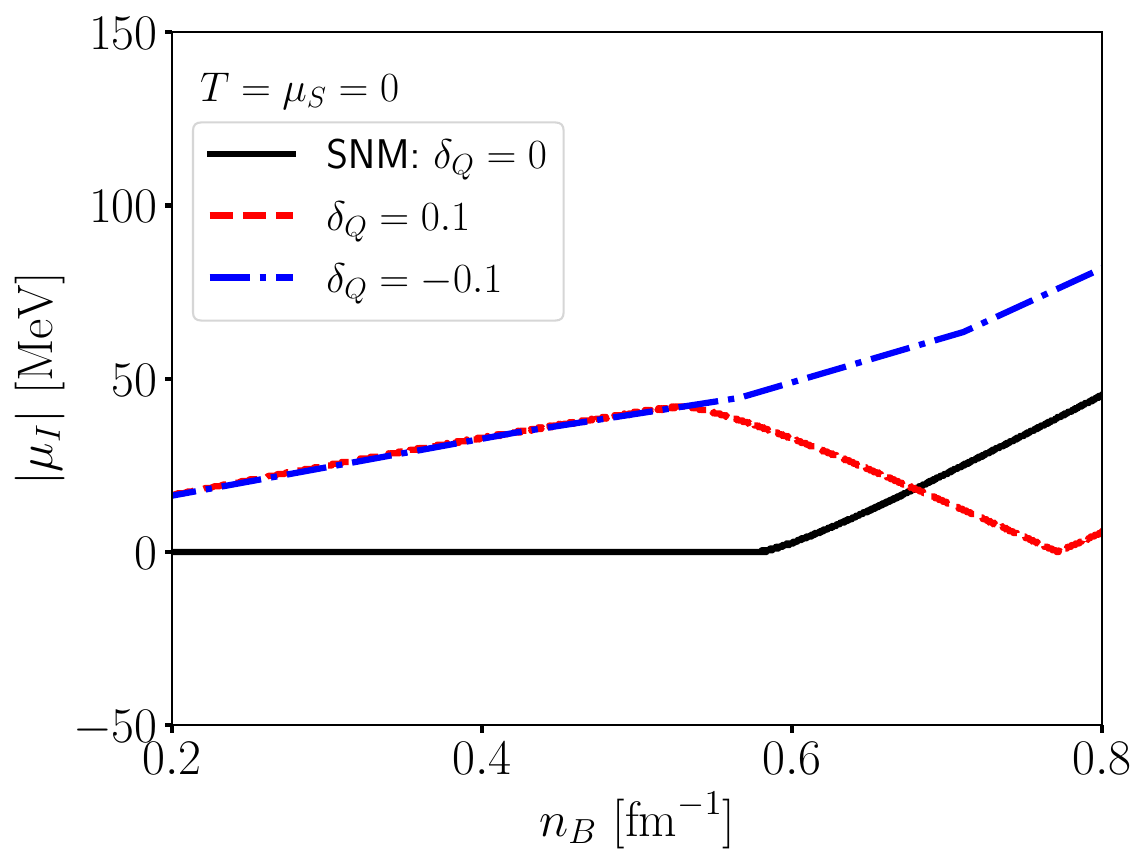} \\
    \includegraphics[width=\linewidth]{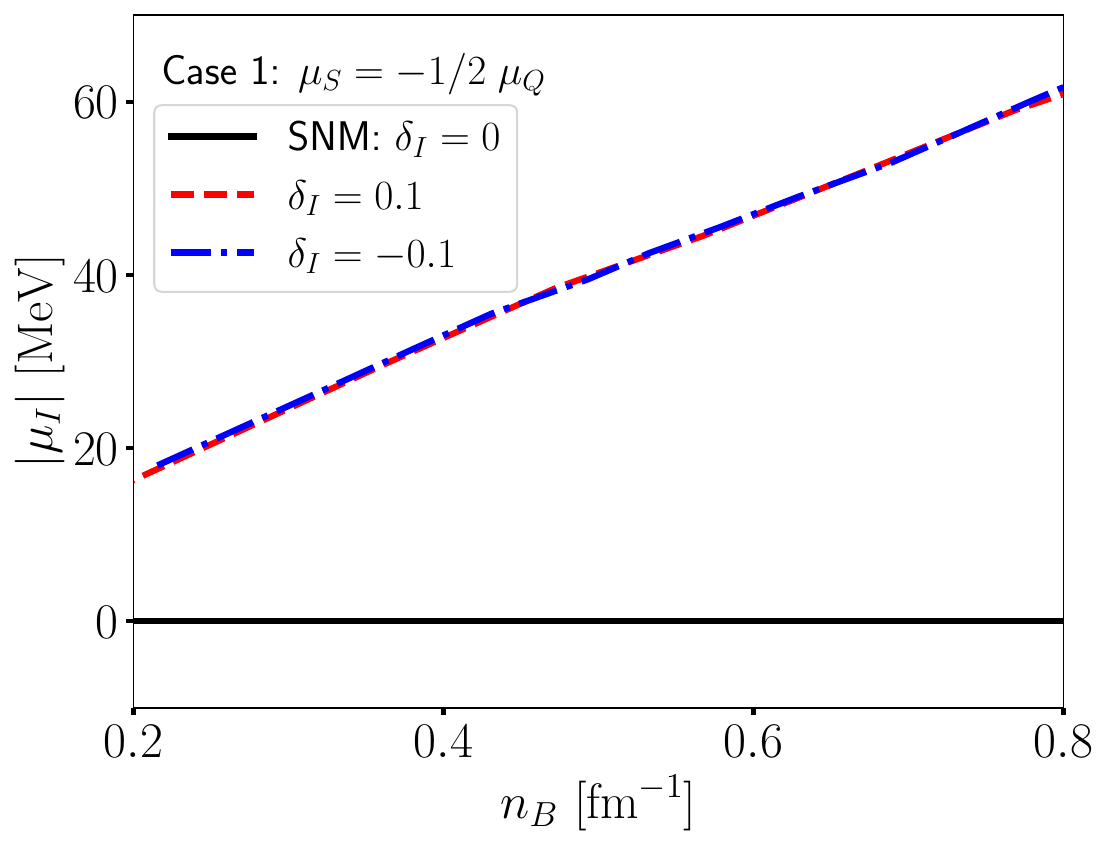}\\
    \includegraphics[width=\linewidth]{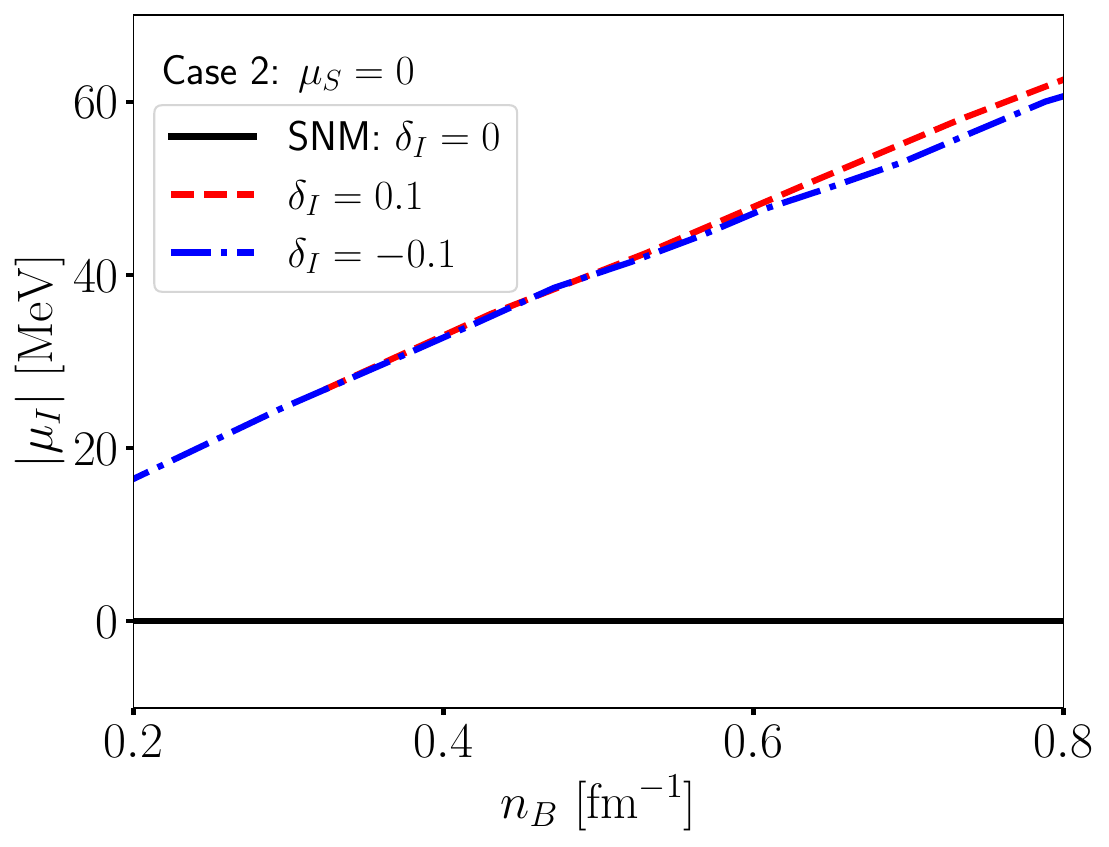}
    \caption{The isospin chemical potential $\mu_I(n_B,\delta)$ for positive and negative values of the asymmetry parameter in its original form, $\delta_Q$ (top), in our new $\delta_I$ form for isospin-reflection-symmetric matter $\mu_S=-1/2\mu_Q$ (middle), and in after redefinition, $\delta_I$ for weak equilibrium $\mu_S=0$ (bottom).}
    \label{fig:muQ}
\end{figure}

Another way to look at the isospin reflection symmetry is via $\mu_I(\delta)$, which should be zero for $\delta=0$. 
For ANM, we can check whether $\mu_I(-\delta) = -\mu_I(\delta)$, which is expected from isospin symmetry. 
In Fig.~\ref{fig:muQ} we show $|\mu_I|(\delta)$ for different choices of $\delta_Q$ (top) and $\delta_I$ (middle,bottom).
We find that $\mu_I\neq 0$ for $\delta_Q=0$ at high $n_B$ because of the appearance of strange particles, which implies once again that $\delta_Q$ is a poor basis for the expansion. 
We then consider two cases for the $\delta_I$ term comparison:
\begin{enumerate}
    \item Isospin-symmetric expansion in $\delta_I$. In Eq.~(\ref{eqn:muIbasis}) we can obtain a reflection in isospin by setting $\mu_S=-\mu_Q/2$ such that, for each fixed $n_B$, the energy change is the same as $I_{z,i} \to - I_{z,i}$.
    \item  Strange break of isospin reflection symmetry in $\delta_I$. In the case of weak equilibrium, $\mu_S=0$, isospin reflection is broken by the contribution of a $(1/2)\mathcal{S}\mu_Q$ term as shown by Eq.~\eqref{eqn:muIbasis}. 
\end{enumerate}
For case 1 where $\mu_S=-\mu_Q/2$ we find perfect isospin reflection of $|\mu_I|(+\delta_I)=|\mu_I|(-\delta_I)$ even after the onset of strangeness ($n_B\gtrsim 0.57\,\mathrm{fm}^{-3}$ for SNM). 
For case 2 where $\mu_S=0$ we find a significant improvement when using $\delta_I$ (compared to $\delta_Q$) such that there is perfect agreement before strangeness switches on, but after $n_S\neq 0$ then we find $|\mu_I|(\delta_I)>|\mu_I|(-\delta_I)$. 
This relationship makes sense given the contribution of the  $(1/2)\mathcal{S}\mu_Q$ that contributes positively for $\delta_I>0$ ($\mathcal{S}<0$ and $\mu_Q<0$) and negatively for $\delta_I>0$ ($\mathcal{S}<0$ but $\mu_Q>0$). 

Before delving into the details of these two cases, we want to introduce the symmetry-energy expansion. 
Originally, the binding energy across a range of nuclei suggested that $\tilde{E}(\delta_Q) = \tilde{E}(-\delta_Q)$  \cite{Bombaci:1991zz}, which was leveraged to eliminate odd terms in $\delta_Q$ and $\delta_Q^3$. 
For just protons and neutrons, $\delta_Q\to-\delta_Q$ symmetry is a good approximation (with certain caveats \cite{Roca-Maza:2018bpv}). 
Theoretically, this parity observation can be understood as a consequence of the isospin reflection symmetry, so we may expect $\tilde{E}(\delta_I)=\tilde{E}(-\delta_I)$ even after the onset of strangeness. 
One expands the energy per nucleon $\tilde{E}/A$ of ANM at some value of $\delta$ around that of SNM, in powers of the asymmetry parameter $\delta$ (here either $\delta_Q$ or $\delta_I$):
\begin{subequations}
\begin{eqnarray}
\frac{\tilde{E}_{ANM}}{A}(n_B,\delta)&=&\frac{\tilde{E}_{SNM}}{A}(n_B)+S(n_B,\delta),\\
S(n_B,\delta) &=&\tilde{E}_{sym,1}(n_B)\,\delta +\tilde{E}_{sym,2}(n_B)\,\delta^2  \nonumber\\ 
    &+&\tilde{E}_{sym,3}(n_B)\,\delta^3+\mathcal{O}(\delta^4), \label{eqn:deltaEXPAN}
\end{eqnarray}
\end{subequations}
where $S(n_B)$ is defined as the difference between the ANM and SNM energies
and we subtract the nucleon rest mass $m_{np}=(m_n+m_p)/2$ from $\tilde{E}/A = \varepsilon/n_B - m_{np}$, where $\varepsilon$ is the energy density.

\begin{figure}[h!]
    \centering
    \includegraphics[width=\linewidth]{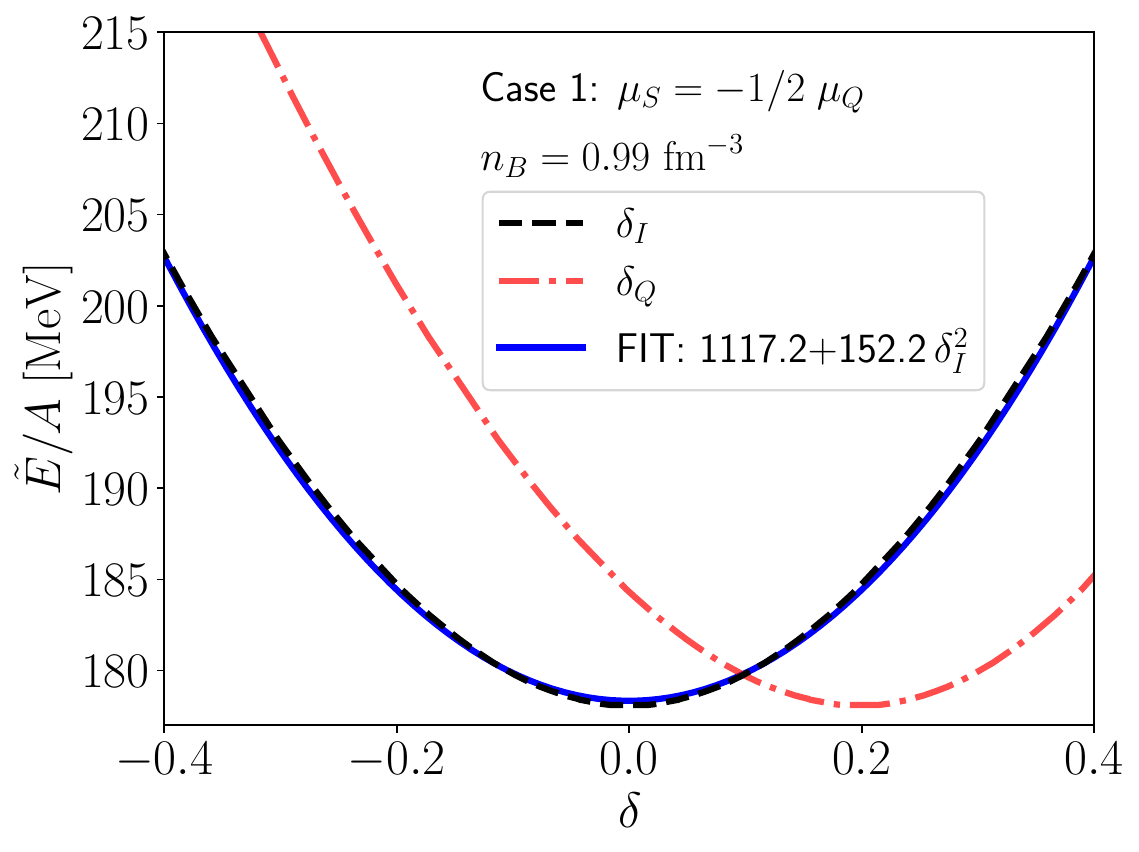}
    \caption{Energy per baryon number vs $\delta$ at a fixed baryon density $n_B=0.99\,\mathrm{fm}^{-3}$. 
    The original $\delta_Q$ finds that the ground state of nuclear matter is at a finite $\delta_Q\approx 0.2$, whereas our new $\delta_I$ has the minimum coincide with $\delta_I=0$. Thus, all linear terms should disappear from the symmetry-energy expansion. The binding energy is symmetric across $\delta_I$. }
    \label{fig:EA_muI}
\end{figure}
\subsection{Case 1: Strangeness isospin reflection symmetry}
We now consider the case of isospin-reflection-symmetric matter that has a reflection symmetry across the SNM axis. 
To do so, we take $\mu_S=-1/2\mu_Q$ within CMF and study the behavior of $\tilde{E}/A$ across either $\delta_Q$ or $\delta_I$, as shown in Fig.\ \ref{fig:EA_muI} at $n_B=0.99\,\mathrm{fm}^{-3}$, which is well above the onset of strangeness.
We find that $\delta_Q$ leads to a significant problem in that the ground state of nuclear matter shifts to a finite $\delta_Q\approx 0.2$ value such that one would obtain linear terms in the symmetry-energy expansion, i.e., $\tilde{E}_{sym,1}\neq 0$. 
However, if we instead use $\delta_I$ it is clear that we have isospin reflection symmetry across the $\delta_I=0$ axis and that an expansion up to $\delta_I^2$ leads to a nearly perfect reproduction of the thermodynamics. 

Thus, for the isospin reflection case for strangeness, i.e., when $\mu_S=-1/2\mu_Q$, our symmetry-energy expansion can be written as:
\begin{equation}
    \frac{\tilde{E}_{ANM}}{A}(n_B,\delta_I)\big|_{\mu_S=-\frac{\mu_Q}{2}}=\frac{\tilde{E}_{SNM}}{A}(n_B)+\tilde{E}_{sym,2}(n_B)\,\delta_I^2, 
\end{equation}
where in the Supplemental Material~\cite{supplemental} we demonstrate a surprisingly accurate description of $\tilde{E}_{sym,2}(n_B)$ expanding around $n_{sat}$ up to $\mathcal{O}(n_B^2)$.

\subsection{Case 2: Strange breaking of isospin reflection symmetry}
In the case of $\beta$-equilibrium (otherwise known as weak equilibrium), both decays are in equilibrium:
\begin{eqnarray}
    d&\rightarrow &u+e^-+\bar{\nu}_e\\
    s&\rightarrow &u+e^-+\bar{\nu}_e
\end{eqnarray}
such that the chemical potentials must hold in the free-streaming neutrino case:
\begin{eqnarray}
    \tilde{\mu}_d&=&\tilde{\mu}_u+\tilde{\mu}_e \\
    \tilde{\mu}_s&=&\tilde{\mu}_u+\tilde{\mu}_e \\
    \therefore \tilde{\mu}_d&=&\tilde{\mu}_s\\
    \therefore \mu_S&=&0
\end{eqnarray}
which we anticipate within old, isolated neutron stars. 
Here, we do not anticipate perfect isospin reflection symmetry, but we expect that the choice of $\delta_I$ will still significantly improve the symmetry-energy expansion. 

\begin{figure}
    \begin{tabular}{cc}
      \includegraphics[width=\linewidth]{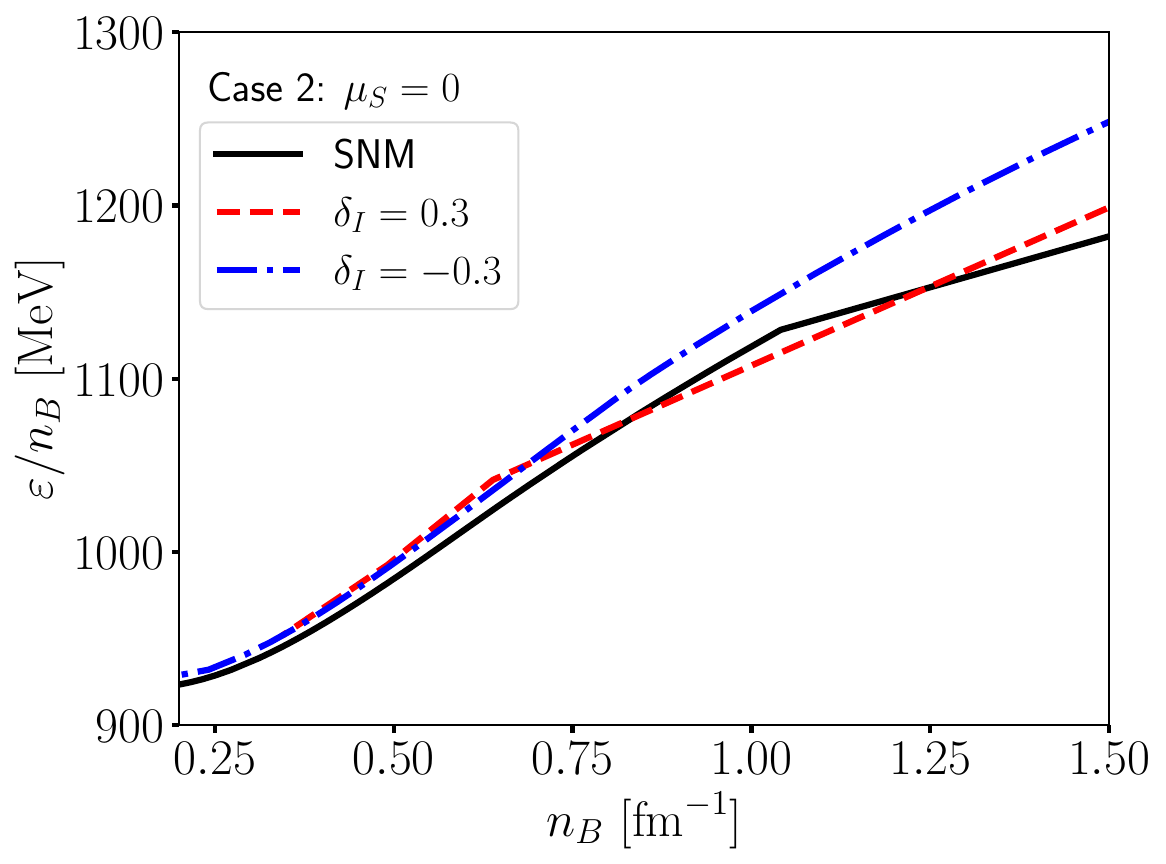}
    \end{tabular}
    \caption{Energy density over baryon density vs baryon density for symmetric nuclear matter $\delta_I=0$, and demonstrating reflection symmetry across the $\delta_I$ axis by comparing  $\delta_I=0.3$ vs $\delta_I=-0.3$.
    At low $n_B$, $Y_S=0$ such that we see a reflection symmetry at finite $\delta_I$ but at $n_B \approx 0.57\,\mathrm{fm}^{-3}$, the reflection symmetry starts to break.
    }
    \label{fig:esym_v_deltas}
\end{figure}

In Fig.\ \ref{fig:esym_v_deltas} we show the energy per baryon, $\varepsilon/n_B(n_B)$, for SNM and $\delta_I=\pm 0.3$. 
We find a perfect symmetry under $\delta_I\to -\delta_I$ at low $n_B$ and a slight breaking of this symmetry above $n_B\gtrsim
0.57\,\mathrm{fm}^{-3}$, but we see an even more pronounced breaking  for $n_B\gtrsim
0.6\,\mathrm{fm}^{-3}$, with $\varepsilon/n_B(\delta_I)>\varepsilon/n_B(-\delta_I)$, which is consistent with the effect in $|\mu_I|(\pm\delta_I)$.

\begin{figure}
    \centering
    \includegraphics[width=\linewidth]{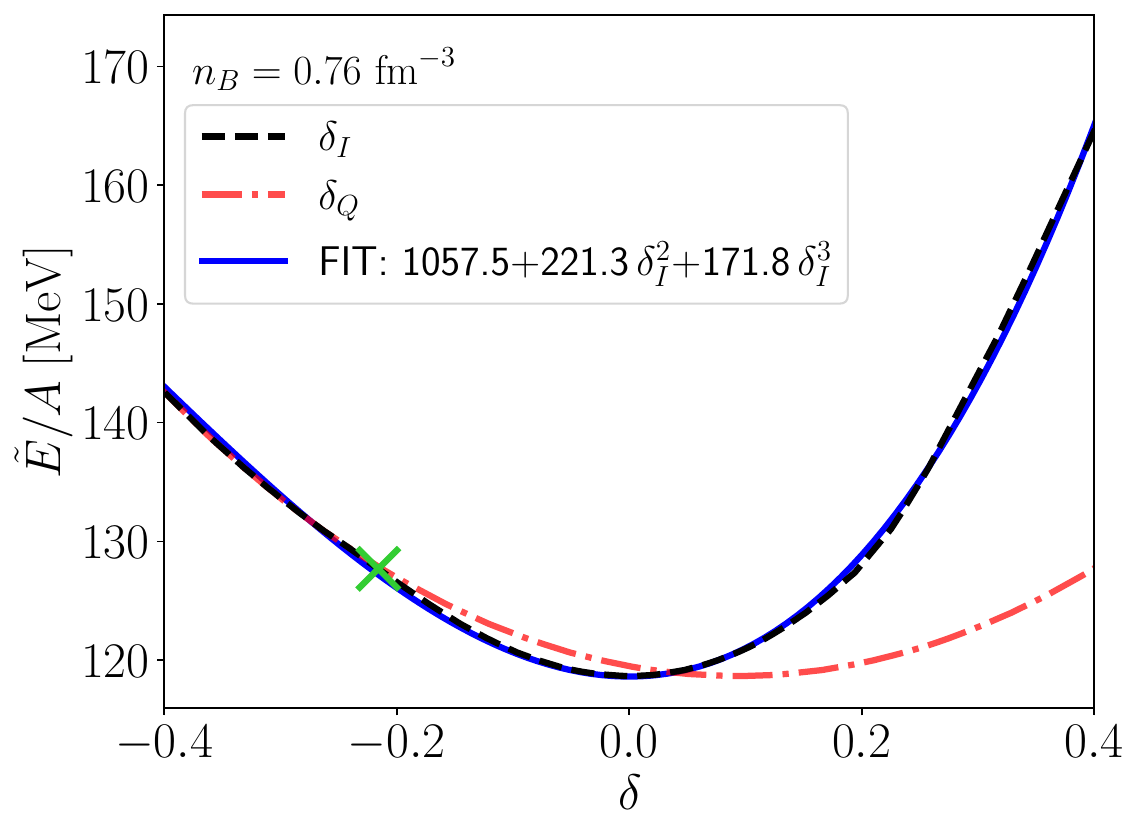}
    \caption{Energy per baryon number vs $\delta_I$ at a fixed baryon density $n_B=0.76\,\mathrm{fm}^{-3}$.  
    The original $\delta_Q$ finds that the ground state of nuclear matter is at a finite $\delta_Q\approx 0.1$, whereas our new $\delta_I$ has the minimum coincide with $\delta_I=0$. Thus, all linear terms should disappear from the symmetry-energy expansion. We can see that the binding energy is asymmetric across $\delta_I$ such that a skewness term ($\delta_I^3$) provides a nearly perfect fit. The green cross marks the onset of strangeness. On the left, $Y_S = 0$, while on the right, $Y_S\neq 0$. 
    }
    \label{fig:skewness}
\end{figure}

\begin{figure}
\includegraphics[width=\linewidth]{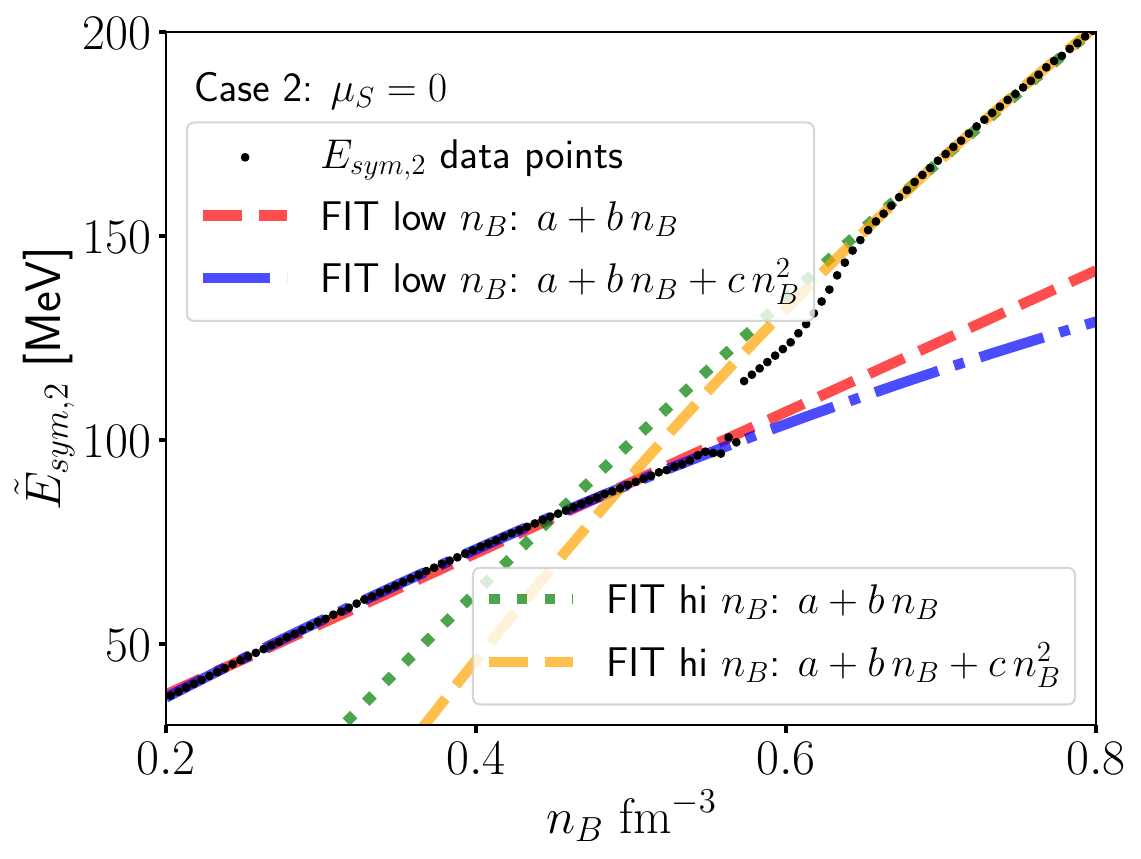}
\includegraphics[width=\linewidth]{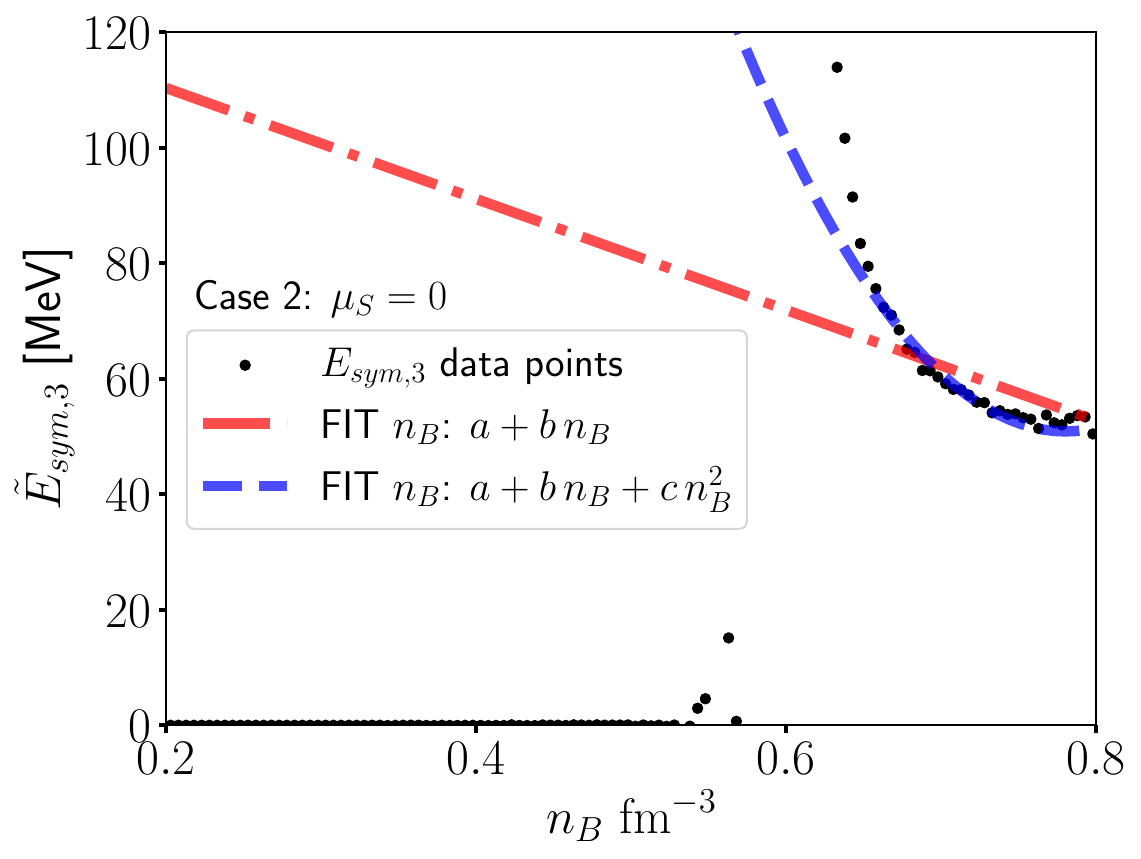}
\caption{Extracted $E_{\textrm{sym},2}(n_B)$ and $E_{\textrm{sym},3}(n_B)$ from a fit assuming quadratic and cubic terms in the symmetry-energy expansion. Fits of different orders are shown to determine the number of coefficients required to reproduce the CMF++ EOS.}
\label{fig:fit_order_CMF}
\end{figure}

In Fig.\ \ref{fig:skewness} we show $\tilde{E}/A(n_B)$  and fit a polynomial in $\delta_I$ or $\delta_Q$ to the data.
The isospin reflection symmetry is broken due to finite $Y_S$, as expected. 
Thus, when $Y_S\neq 0$: 
\begin{enumerate}
    \item Using $\delta_Q$ shifts the ground state of nuclear matter to a finite $\delta_Q\approx 0.03$, breaking linear symmetry.
    \item Using $\delta_I$ shifts the ground state of nuclear matter back to $\delta_I=0$, restoring linear symmetry. 
    \item Weak equilibrium ($\mu_S=0$) leads to a skewness across $\delta_I$ such that $E_{\textrm{sym},3}\neq 0$ in Eq.~\eqref{eqn:deltaEXPAN}. The skewness term, when it becomes nonzero, is of a comparable magnitude to the traditional symmetry energy $E_{\mathrm{sym},2}$, as shown by Fig.~\ref{fig:fit_order_CMF}.
\end{enumerate}

\begin{figure}
    \begin{tabular}{cc}

    \includegraphics[width=\linewidth]{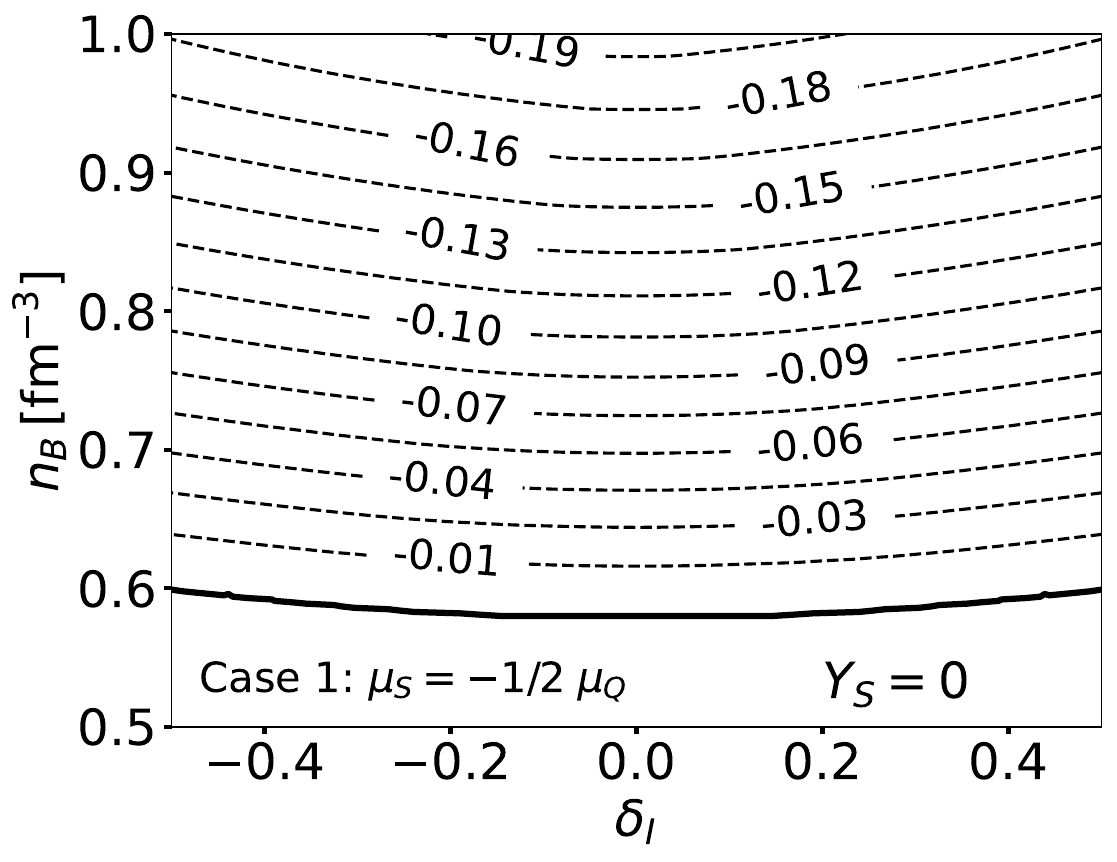} \\
    \includegraphics[width=\linewidth]{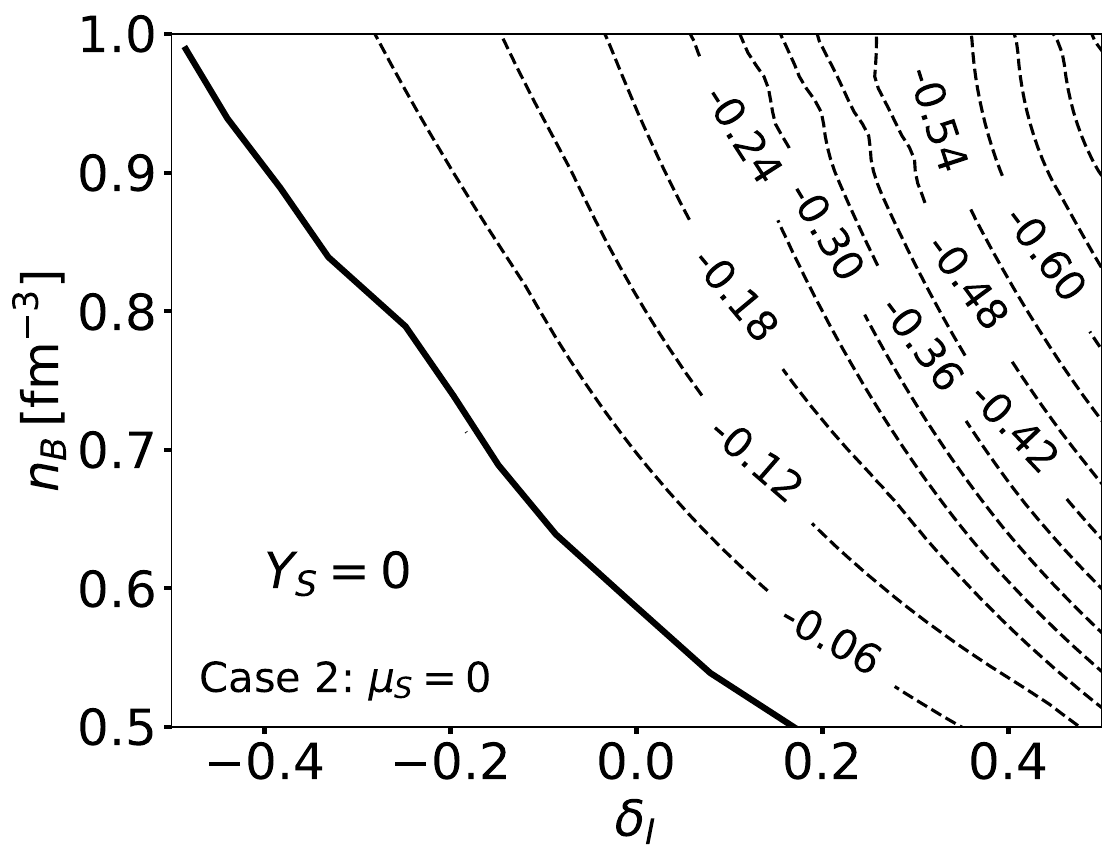}
    \end{tabular}
    \caption{Contour plot of constant strangeness fraction $Y_S$ projected into the plane of baryon density vs. $\delta_I$, with isospin-reflection-symmetric system (top), and system in weak equilibrium (bottom).}
    \label{fig:nB_vs_deltaS_contour}
\end{figure}

\section{Phase Diagram of $Y_S$}
One of the most crucial differences in our two cases listed above is the phase diagram of $Y_S(n_B,\delta_I)$.
In case 2 of weak equilibrium, $\mu_S=0$ such that $\mu_S^{(B,I_z)}=\frac{1}{2}\mu_Q$. Positive $\delta_I$ leads to $\mu_I<0$ (recall in this limit $\mu_Q=\mu_I$) such that $\mu_S^{(B,I_z)}<0$, and since hyperons carry negative strangeness, it is easier to produce strange particles. In contrast,  $\delta_I < 0$ suppresses strangeness. 
Once $Y_S\neq 0$, this leads to asymmetries under $\delta_I \to -\delta_I$. 
In contrast, in case 1, where $\mu_S^{(B,I_z)}=0$, there is an equal probability to produce strangeness in both the $\pm\mu_I$ directions, such that $Y_S$ is symmetric around the $\delta_I$ axis.  

In Fig.~\ref{fig:nB_vs_deltaS_contour}, we can validate these arguments where case 1 is shown (top) to have $Y_S$ that is, indeed, symmetric around the $\delta_I$ axis. 
In contrast we find that $Y_S$ has a clear skewness across the $\delta_I$ axis because it is easier to produce strangeness for $\delta_I>0$ whereas strangeness is suppressed for $\delta_I<0$. 
This skewness in strangeness for the weak equilibrium case is precisely what leads to the appearance of the symmetry-breaking strangeness term $E_{\textrm{sym},3}\neq 0$.

\section{Conclusions and Outlook}
In this paper, we found that the typical isospin-asymmetry coefficient $\delta_Q=1-2Y_Q$ does not obtain the ground state of nuclear matter when strangeness is finite. 
Thus, we have rewritten the isospin-asymmetry coefficient $\delta_I$ to be inclusive of strangeness, which can obtain the ground state of nuclear matter for symmetric nuclear matter both with and without strangeness.
We then use CMF to consider two cases: one that preserves isospin reflection symmetry in the presence of strangeness $\mu_S=-1/2\mu_Q$ and the other case associated with old, cold neutron stars of weak equilibrium $\mu_S=0$ that breaks isospin reflection symmetry. 

While previous works have investigated correlations between symmetry-energy coefficients and strangeness content and other quantities \cite{Providencia:2012rx, Providencia:2013dsa},  or
at most considered the asymmetry between protons and neutrons in the presence of hyperons \cite{Bednarek:2019xytNI}, we have extended the symmetry-energy expansion to strange matter via general considerations regarding the symmetries of the strong and electroweak interactions, without making any extra assumptions on particle content besides the presence of strange particles.
As a result, our considerations apply both to nuclear and quark matter alike \cite{Danhoni:2025qpn}.
This is also the first time an isospin-breaking skewness term is included in the symmetry-energy expansion. 
While Figs.~\ref{fig:EA_muI} and \ref{fig:skewness} show that our generalized expansion accurately reproduces the energy per baryon, the conventional symmetry-energy expansion can deviate substantially. For instance, at $n_B = 0.7\,\mathrm{fm}^{-3}$ the conventional expansion underestimates $\tilde{E}/A$ by $\approx 80\%$ relative to the underlying EOS.

However, although our new symmetry-energy expansion holds well beyond typical neutron star maximum central densities, it fails eventually at higher densities. For isospin-reflection-symmetric matter, i.e., $\mu_S=-1/2\mu_Q$ (in the basis where odd terms are symmetry forbidden), the expansion around $n_{sat}$ remains accurate through densities exceeding $\approx 7\,n_{sat}$ within the tested range of $\delta_I$. For strange matter in weak equilibrium ($\mu_S=0$), we find that the truncated Taylor-like description begins to fail at around $\approx 5.5\,n_{\rm sat}$, in the sense that $E/A$ as a function of $\delta_I$ develops additional structure beyond what can be captured by a low-order polynomial truncation about a single expansion point. In addition, our expansion assumes a single homogeneous thermodynamic phase and sufficient smoothness (analyticity) of the relevant thermodynamic potential in $\delta_I$ in the neighborhood of the expansion point(s). If the underlying EOS exhibits a first-order transition (or an extended mixed-phase region) in the strangeness-bearing regime, nonanalytic behavior can invalidate a single global Taylor-like truncation; in such cases, one should treat each phase separately (or adopt a phase-coexistence construction) and match across the transition (or see \cite{Parotto:2018pwx, Kahangirwe:2024cny} for an alternative method).

Our results have led to two very interesting findings that can pave the way for future work. 
The first finding is that $Y_S\neq0$ leads to a skewness around the ground state of nuclear matter in the case of weak equilibrium (given a specific $n_B$). In other words, strangeness implies isospin reflection symmetry breaking across the SNM axis when $\mu_S=0$. 
We have also checked our framework in an ideal free Fermi gas including the full baryon octet and decuplet, and we performed an independent perturbative QCD (pQCD) analysis in Ref.~\cite{Danhoni:2025qpn}. In the pQCD case, the extracted expansion coefficients are of comparable order of magnitude.
It is likely possible to mathematically connect the isospin-reflection-symmetric system of $\mu_S=-1/2\mu_Q$ to that of weak equilibrium, which we leave for a future work.
However, we believe that this strangeness breaking of isospin reflection symmetry may help to explain some of the interesting phase transitions that appear at $T=0$ across finite $\mu_S$ and $\mu_Q$ in CMF++ \cite{Cruz-Camacho:2024odu}. 

The second implication of our findings is that new methods are needed to extract symmetry-energy coefficients in the presence of strangeness from experimental data,  chiral effective field theory, and other theoretical models. 
One potential possibility is creating systems that include finite strangeness. 
There are two possible types of experiments that we envision that could provide information about strangeness. The first would be fixed-target collisions of kaon beams (see, e.g., Ref.~\cite{NA62KLEVER:2022nea} for other experiments with kaon beams) with a nucleus and the second would be collisions of hypernuclei. Similarly, it may be possible to obtain information about these coefficients in chiral effective field theory if strange interactions were included (see, e.g., Refs.~\cite{Haidenbauer:2009qn,Li:2018tbt,Song:2018qqm,Haidenbauer:2015zqb}). 
Finally, varying the large parameter space of CMF++ (and other similar EOS models that include hyperons), while constraining it to reproduce saturation properties and neutron star properties \cite{MUSES:2023hyz}, can help to extract the allowable range of these symmetry coefficients. 

Moreover, the strange symmetry-energy expansion provides a compact set of density-dependent functions
$\{\tilde E_{\rm sym,k}(n_B)\}$ (and, when needed, their low-density and strangeness-onset expansions)
that can serve as a convenient parametrization of the cold nuclear-matter EOS in the presence of finite
strangeness. This is particularly well suited for Bayesian inference frameworks that seek to constrain the
neutron star EOS from multimessenger observations, since the expansion coefficients can be sampled directly
as low-dimensional parameters (with priors informed by nuclear theory and laboratory data) and then mapped
to posterior distributions for neutron star observables.
At the same time, the same coefficients control the EOS-dependent part of weak-interaction-driven flavor
equilibration, and therefore enter the microphysical inputs for dissipative hydrodynamics in mergers, such
as bulk viscosity and the associated relaxation timescales \cite{Yang:2025yoo, Harris:2025ncu}. A natural next step is
to propagate EOS posteriors for $\{\tilde E_{\rm sym,k}\}$ into posteriors for transport coefficients in merger
simulations, thereby enabling a consistent quantification of both equilibrium and out-of-equilibrium uncertainties.

\begin{acknowledgments}
This research was
partly supported by the U.S. DOE Nuclear Science Grant No. DE-SC0023861 and by
the National Science Foundation (NSF) within the framework of the MUSES Collaboration, under Grant No.
OAC-2103680. Any opinions, findings, and conclusions
or recommendations expressed in this material are those
of the author(s) and do not necessarily reflect the views of
the National Science Foundation. We also acknowledge
support from the Illinois Campus Cluster, a computing
resource that is operated by the Illinois Campus Cluster Program (ICCP) in conjunction with the National
Center for Supercomputing Applications (NCSA), which
is supported by funds from the University of Illinois at
Urbana-Champaign. 
M.H. was supported by the Brazilian National Council for Scientific and Technological Development (CNPq) under Process No. 313638/2025-0. 
\end{acknowledgments}

\section*{Data Availability}
The data that support the findings of this article are openly available~\cite{Cruz-Camacho:2024odu}.

\bibliography{inspire,NOTinspire}

\begin{thebibliography}{95}%
\makeatletter
\providecommand \@ifxundefined [1]{%
 \@ifx{#1\undefined}
}%
\providecommand \@ifnum [1]{%
 \ifnum #1\expandafter \@firstoftwo
 \else \expandafter \@secondoftwo
 \fi
}%
\providecommand \@ifx [1]{%
 \ifx #1\expandafter \@firstoftwo
 \else \expandafter \@secondoftwo
 \fi
}%
\providecommand \natexlab [1]{#1}%
\providecommand \enquote  [1]{``#1''}%
\providecommand \bibnamefont  [1]{#1}%
\providecommand \bibfnamefont [1]{#1}%
\providecommand \citenamefont [1]{#1}%
\providecommand \href@noop [0]{\@secondoftwo}%
\providecommand \href [0]{\begingroup \@sanitize@url \@href}%
\providecommand \@href[1]{\@@startlink{#1}\@@href}%
\providecommand \@@href[1]{\endgroup#1\@@endlink}%
\providecommand \@sanitize@url [0]{\catcode `\\12\catcode `\$12\catcode
  `\&12\catcode `\#12\catcode `\^12\catcode `\_12\catcode `\%12\relax}%
\providecommand \@@startlink[1]{}%
\providecommand \@@endlink[0]{}%
\providecommand \url  [0]{\begingroup\@sanitize@url \@url }%
\providecommand \@url [1]{\endgroup\@href {#1}{\urlprefix }}%
\providecommand \urlprefix  [0]{URL }%
\providecommand \Eprint [0]{\href }%
\providecommand \doibase [0]{http://dx.doi.org/}%
\providecommand \selectlanguage [0]{\@gobble}%
\providecommand \bibinfo  [0]{\@secondoftwo}%
\providecommand \bibfield  [0]{\@secondoftwo}%
\providecommand \translation [1]{[#1]}%
\providecommand \BibitemOpen [0]{}%
\providecommand \bibitemStop [0]{}%
\providecommand \bibitemNoStop [0]{.\EOS\space}%
\providecommand \EOS [0]{\spacefactor3000\relax}%
\providecommand \BibitemShut  [1]{\csname bibitem#1\endcsname}%
\let\auto@bib@innerbib\@empty
\bibitem [{\citenamefont {Collins}\ and\ \citenamefont
  {Perry}(1975)}]{Collins:1974ky}%
  \BibitemOpen
  \bibfield  {author} {\bibinfo {author} {\bibfnamefont {J.~C.}\ \bibnamefont
  {Collins}}\ and\ \bibinfo {author} {\bibfnamefont {M.~J.}\ \bibnamefont
  {Perry}},\ }\href {\doibase 10.1103/PhysRevLett.34.1353} {\bibfield
  {journal} {\bibinfo  {journal} {Phys. Rev. Lett.}\ }\textbf {\bibinfo
  {volume} {34}},\ \bibinfo {pages} {1353} (\bibinfo {year}
  {1975})}\BibitemShut {NoStop}%
\bibitem [{\citenamefont {Cabibbo}\ and\ \citenamefont
  {Parisi}(1975)}]{Cabibbo:1975ig}%
  \BibitemOpen
  \bibfield  {author} {\bibinfo {author} {\bibfnamefont {N.}~\bibnamefont
  {Cabibbo}}\ and\ \bibinfo {author} {\bibfnamefont {G.}~\bibnamefont
  {Parisi}},\ }\href {\doibase 10.1016/0370-2693(75)90158-6} {\bibfield
  {journal} {\bibinfo  {journal} {Phys. Lett. B}\ }\textbf {\bibinfo {volume}
  {59}},\ \bibinfo {pages} {67} (\bibinfo {year} {1975})}\BibitemShut {NoStop}%
\bibitem [{\citenamefont {Bzdak}\ \emph {et~al.}(2020)\citenamefont {Bzdak},
  \citenamefont {Esumi}, \citenamefont {Koch}, \citenamefont {Liao},
  \citenamefont {Stephanov},\ and\ \citenamefont {Xu}}]{Bzdak:2019pkr}%
  \BibitemOpen
  \bibfield  {author} {\bibinfo {author} {\bibfnamefont {A.}~\bibnamefont
  {Bzdak}}, \bibinfo {author} {\bibfnamefont {S.}~\bibnamefont {Esumi}},
  \bibinfo {author} {\bibfnamefont {V.}~\bibnamefont {Koch}}, \bibinfo {author}
  {\bibfnamefont {J.}~\bibnamefont {Liao}}, \bibinfo {author} {\bibfnamefont
  {M.}~\bibnamefont {Stephanov}}, \ and\ \bibinfo {author} {\bibfnamefont
  {N.}~\bibnamefont {Xu}},\ }\href {\doibase 10.1016/j.physrep.2020.01.005}
  {\bibfield  {journal} {\bibinfo  {journal} {Phys. Rept.}\ }\textbf {\bibinfo
  {volume} {853}},\ \bibinfo {pages} {1} (\bibinfo {year} {2020})},\ \Eprint
  {http://arxiv.org/abs/1906.00936} {arXiv:1906.00936 [nucl-th]} \BibitemShut
  {NoStop}%
\bibitem [{\citenamefont {An}\ \emph {et~al.}(2022)\citenamefont {An} \emph
  {et~al.}}]{An:2021wof}%
  \BibitemOpen
  \bibfield  {author} {\bibinfo {author} {\bibfnamefont {X.}~\bibnamefont {An}}
  \emph {et~al.},\ }\href {\doibase 10.1016/j.nuclphysa.2021.122343} {\bibfield
   {journal} {\bibinfo  {journal} {Nucl. Phys. A}\ }\textbf {\bibinfo {volume}
  {1017}},\ \bibinfo {pages} {122343} (\bibinfo {year} {2022})},\ \Eprint
  {http://arxiv.org/abs/2108.13867} {arXiv:2108.13867 [nucl-th]} \BibitemShut
  {NoStop}%
\bibitem [{\citenamefont {Sorensen}\ \emph {et~al.}(2024)\citenamefont
  {Sorensen} \emph {et~al.}}]{Sorensen:2023zkk}%
  \BibitemOpen
  \bibfield  {author} {\bibinfo {author} {\bibfnamefont {A.}~\bibnamefont
  {Sorensen}} \emph {et~al.},\ }\href {\doibase 10.1016/j.ppnp.2023.104080}
  {\bibfield  {journal} {\bibinfo  {journal} {Prog. Part. Nucl. Phys.}\
  }\textbf {\bibinfo {volume} {134}},\ \bibinfo {pages} {104080} (\bibinfo
  {year} {2024})},\ \Eprint {http://arxiv.org/abs/2301.13253} {arXiv:2301.13253
  [nucl-th]} \BibitemShut {NoStop}%
\bibitem [{\citenamefont {Lovato}\ \emph {et~al.}(2022)\citenamefont {Lovato}
  \emph {et~al.}}]{Lovato:2022vgq}%
  \BibitemOpen
  \bibfield  {author} {\bibinfo {author} {\bibfnamefont {A.}~\bibnamefont
  {Lovato}} \emph {et~al.},\ }\href@noop {} {\  (\bibinfo {year} {2022})},\
  \Eprint {http://arxiv.org/abs/2211.02224} {arXiv:2211.02224 [nucl-th]}
  \BibitemShut {NoStop}%
\bibitem [{\citenamefont {Danielewicz}\ \emph {et~al.}(2002)\citenamefont
  {Danielewicz}, \citenamefont {Lacey},\ and\ \citenamefont
  {Lynch}}]{Danielewicz:2002pu}%
  \BibitemOpen
  \bibfield  {author} {\bibinfo {author} {\bibfnamefont {P.}~\bibnamefont
  {Danielewicz}}, \bibinfo {author} {\bibfnamefont {R.}~\bibnamefont {Lacey}},
  \ and\ \bibinfo {author} {\bibfnamefont {W.~G.}\ \bibnamefont {Lynch}},\
  }\href {\doibase 10.1126/science.1078070} {\bibfield  {journal} {\bibinfo
  {journal} {Science}\ }\textbf {\bibinfo {volume} {298}},\ \bibinfo {pages}
  {1592} (\bibinfo {year} {2002})},\ \Eprint
  {http://arxiv.org/abs/nucl-th/0208016} {arXiv:nucl-th/0208016} \BibitemShut
  {NoStop}%
\bibitem [{\citenamefont {Pratt}\ \emph {et~al.}(2015)\citenamefont {Pratt},
  \citenamefont {Sangaline}, \citenamefont {Sorensen},\ and\ \citenamefont
  {Wang}}]{Pratt:2015zsa}%
  \BibitemOpen
  \bibfield  {author} {\bibinfo {author} {\bibfnamefont {S.}~\bibnamefont
  {Pratt}}, \bibinfo {author} {\bibfnamefont {E.}~\bibnamefont {Sangaline}},
  \bibinfo {author} {\bibfnamefont {P.}~\bibnamefont {Sorensen}}, \ and\
  \bibinfo {author} {\bibfnamefont {H.}~\bibnamefont {Wang}},\ }\href {\doibase
  10.1103/PhysRevLett.114.202301} {\bibfield  {journal} {\bibinfo  {journal}
  {Phys. Rev. Lett.}\ }\textbf {\bibinfo {volume} {114}},\ \bibinfo {pages}
  {202301} (\bibinfo {year} {2015})},\ \Eprint
  {http://arxiv.org/abs/1501.04042} {arXiv:1501.04042 [nucl-th]} \BibitemShut
  {NoStop}%
\bibitem [{\citenamefont {Ablyazimov}\ \emph {et~al.}(2017)\citenamefont
  {Ablyazimov} \emph {et~al.}}]{CBM:2016kpk}%
  \BibitemOpen
  \bibfield  {author} {\bibinfo {author} {\bibfnamefont {T.}~\bibnamefont
  {Ablyazimov}} \emph {et~al.} (\bibinfo {collaboration} {CBM}),\ }\href
  {\doibase 10.1140/epja/i2017-12248-y} {\bibfield  {journal} {\bibinfo
  {journal} {Eur. Phys. J. A}\ }\textbf {\bibinfo {volume} {53}},\ \bibinfo
  {pages} {60} (\bibinfo {year} {2017})},\ \Eprint
  {http://arxiv.org/abs/1607.01487} {arXiv:1607.01487 [nucl-ex]} \BibitemShut
  {NoStop}%
\bibitem [{\citenamefont {Alba}\ \emph {et~al.}(2018)\citenamefont {Alba},
  \citenamefont {Mantovani~Sarti}, \citenamefont {Noronha}, \citenamefont
  {Noronha-Hostler}, \citenamefont {Parotto}, \citenamefont
  {Portillo~Vazquez},\ and\ \citenamefont {Ratti}}]{Alba:2017hhe}%
  \BibitemOpen
  \bibfield  {author} {\bibinfo {author} {\bibfnamefont {P.}~\bibnamefont
  {Alba}}, \bibinfo {author} {\bibfnamefont {V.}~\bibnamefont
  {Mantovani~Sarti}}, \bibinfo {author} {\bibfnamefont {J.}~\bibnamefont
  {Noronha}}, \bibinfo {author} {\bibfnamefont {J.}~\bibnamefont
  {Noronha-Hostler}}, \bibinfo {author} {\bibfnamefont {P.}~\bibnamefont
  {Parotto}}, \bibinfo {author} {\bibfnamefont {I.}~\bibnamefont
  {Portillo~Vazquez}}, \ and\ \bibinfo {author} {\bibfnamefont
  {C.}~\bibnamefont {Ratti}},\ }\href {\doibase 10.1103/PhysRevC.98.034909}
  {\bibfield  {journal} {\bibinfo  {journal} {Phys. Rev. C}\ }\textbf {\bibinfo
  {volume} {98}},\ \bibinfo {pages} {034909} (\bibinfo {year} {2018})},\
  \Eprint {http://arxiv.org/abs/1711.05207} {arXiv:1711.05207 [nucl-th]}
  \BibitemShut {NoStop}%
\bibitem [{\citenamefont {Spieles}\ and\ \citenamefont
  {Bleicher}(2020)}]{Spieles:2020zaa}%
  \BibitemOpen
  \bibfield  {author} {\bibinfo {author} {\bibfnamefont {C.}~\bibnamefont
  {Spieles}}\ and\ \bibinfo {author} {\bibfnamefont {M.}~\bibnamefont
  {Bleicher}},\ }\href {\doibase 10.1140/epjst/e2020-000102-4} {\bibfield
  {journal} {\bibinfo  {journal} {Eur. Phys. J. ST}\ }\textbf {\bibinfo
  {volume} {229}},\ \bibinfo {pages} {3537} (\bibinfo {year} {2020})},\ \Eprint
  {http://arxiv.org/abs/2006.01220} {arXiv:2006.01220 [nucl-th]} \BibitemShut
  {NoStop}%
\bibitem [{\citenamefont {Monnai}\ \emph {et~al.}(2021)\citenamefont {Monnai},
  \citenamefont {Schenke},\ and\ \citenamefont {Shen}}]{Monnai:2021kgu}%
  \BibitemOpen
  \bibfield  {author} {\bibinfo {author} {\bibfnamefont {A.}~\bibnamefont
  {Monnai}}, \bibinfo {author} {\bibfnamefont {B.}~\bibnamefont {Schenke}}, \
  and\ \bibinfo {author} {\bibfnamefont {C.}~\bibnamefont {Shen}},\ }\href
  {\doibase 10.1142/S0217751X21300076} {\bibfield  {journal} {\bibinfo
  {journal} {Int. J. Mod. Phys. A}\ }\textbf {\bibinfo {volume} {36}},\
  \bibinfo {pages} {2130007} (\bibinfo {year} {2021})},\ \Eprint
  {http://arxiv.org/abs/2101.11591} {arXiv:2101.11591 [nucl-th]} \BibitemShut
  {NoStop}%
\bibitem [{\citenamefont {Oliinychenko}\ \emph {et~al.}(2023)\citenamefont
  {Oliinychenko}, \citenamefont {Sorensen}, \citenamefont {Koch},\ and\
  \citenamefont {McLerran}}]{Oliinychenko:2022uvy}%
  \BibitemOpen
  \bibfield  {author} {\bibinfo {author} {\bibfnamefont {D.}~\bibnamefont
  {Oliinychenko}}, \bibinfo {author} {\bibfnamefont {A.}~\bibnamefont
  {Sorensen}}, \bibinfo {author} {\bibfnamefont {V.}~\bibnamefont {Koch}}, \
  and\ \bibinfo {author} {\bibfnamefont {L.}~\bibnamefont {McLerran}},\ }\href
  {\doibase 10.1103/PhysRevC.108.034908} {\bibfield  {journal} {\bibinfo
  {journal} {Phys. Rev. C}\ }\textbf {\bibinfo {volume} {108}},\ \bibinfo
  {pages} {034908} (\bibinfo {year} {2023})},\ \Eprint
  {http://arxiv.org/abs/2208.11996} {arXiv:2208.11996 [nucl-th]} \BibitemShut
  {NoStop}%
\bibitem [{\citenamefont {Yao}\ \emph {et~al.}(2024)\citenamefont {Yao},
  \citenamefont {Sorensen}, \citenamefont {Dexheimer},\ and\ \citenamefont
  {Noronha-Hostler}}]{Yao:2023yda}%
  \BibitemOpen
  \bibfield  {author} {\bibinfo {author} {\bibfnamefont {N.}~\bibnamefont
  {Yao}}, \bibinfo {author} {\bibfnamefont {A.}~\bibnamefont {Sorensen}},
  \bibinfo {author} {\bibfnamefont {V.}~\bibnamefont {Dexheimer}}, \ and\
  \bibinfo {author} {\bibfnamefont {J.}~\bibnamefont {Noronha-Hostler}},\
  }\href {\doibase 10.1103/PhysRevC.109.065803} {\bibfield  {journal} {\bibinfo
   {journal} {Phys. Rev. C}\ }\textbf {\bibinfo {volume} {109}},\ \bibinfo
  {pages} {065803} (\bibinfo {year} {2024})},\ \Eprint
  {http://arxiv.org/abs/2311.18819} {arXiv:2311.18819 [nucl-th]} \BibitemShut
  {NoStop}%
\bibitem [{\citenamefont {Plumberg}\ \emph {et~al.}(2025)\citenamefont
  {Plumberg} \emph {et~al.}}]{Plumberg:2024leb}%
  \BibitemOpen
  \bibfield  {author} {\bibinfo {author} {\bibfnamefont {C.}~\bibnamefont
  {Plumberg}} \emph {et~al.},\ }\href {\doibase 10.1103/PhysRevC.111.044905}
  {\bibfield  {journal} {\bibinfo  {journal} {Phys. Rev. C}\ }\textbf {\bibinfo
  {volume} {111}},\ \bibinfo {pages} {044905} (\bibinfo {year} {2025})},\
  \Eprint {http://arxiv.org/abs/2405.09648} {arXiv:2405.09648 [nucl-th]}
  \BibitemShut {NoStop}%
\bibitem [{\citenamefont {Abbott}\ \emph {et~al.}(2017)\citenamefont {Abbott}
  \emph {et~al.}}]{LIGOScientific:2017ync}%
  \BibitemOpen
  \bibfield  {author} {\bibinfo {author} {\bibfnamefont {B.~P.}\ \bibnamefont
  {Abbott}} \emph {et~al.} (\bibinfo {collaboration} {LIGO Scientific, Virgo,
  Fermi GBM, INTEGRAL, IceCube, AstroSat Cadmium Zinc Telluride Imager Team,
  IPN, Insight-Hxmt, ANTARES, Swift, AGILE Team, 1M2H Team, Dark Energy Camera
  GW-EM, DES, DLT40, GRAWITA, Fermi-LAT, ATCA, ASKAP, Las Cumbres Observatory
  Group, OzGrav, DWF (Deeper Wider Faster Program), AST3, CAASTRO, VINROUGE,
  MASTER, J-GEM, GROWTH, JAGWAR, CaltechNRAO, TTU-NRAO, NuSTAR, Pan-STARRS,
  MAXI Team, TZAC Consortium, KU, Nordic Optical Telescope, ePESSTO, GROND,
  Texas Tech University, SALT Group, TOROS, BOOTES, MWA, CALET, IKI-GW
  Follow-up, H.E.S.S., LOFAR, LWA, HAWC, Pierre Auger, ALMA, Euro VLBI Team, Pi
  of Sky, Chandra Team at McGill University, DFN, ATLAS Telescopes, High Time
  Resolution Universe Survey, RIMAS, RATIR, SKA South Africa/MeerKAT}),\ }\href
  {\doibase 10.3847/2041-8213/aa91c9} {\bibfield  {journal} {\bibinfo
  {journal} {Astrophys. J. Lett.}\ }\textbf {\bibinfo {volume} {848}},\
  \bibinfo {pages} {L12} (\bibinfo {year} {2017})},\ \Eprint
  {http://arxiv.org/abs/1710.05833} {arXiv:1710.05833 [astro-ph.HE]}
  \BibitemShut {NoStop}%
\bibitem [{\citenamefont {Miller}\ \emph {et~al.}(2019)\citenamefont {Miller}
  \emph {et~al.}}]{Miller:2019cac}%
  \BibitemOpen
  \bibfield  {author} {\bibinfo {author} {\bibfnamefont {M.~C.}\ \bibnamefont
  {Miller}} \emph {et~al.},\ }\href {\doibase 10.3847/2041-8213/ab50c5}
  {\bibfield  {journal} {\bibinfo  {journal} {Astrophys. J. Lett.}\ }\textbf
  {\bibinfo {volume} {887}},\ \bibinfo {pages} {L24} (\bibinfo {year}
  {2019})},\ \Eprint {http://arxiv.org/abs/1912.05705} {arXiv:1912.05705
  [astro-ph.HE]} \BibitemShut {NoStop}%
\bibitem [{\citenamefont {Miller}\ \emph {et~al.}(2021)\citenamefont {Miller}
  \emph {et~al.}}]{Miller:2021qha}%
  \BibitemOpen
  \bibfield  {author} {\bibinfo {author} {\bibfnamefont {M.~C.}\ \bibnamefont
  {Miller}} \emph {et~al.},\ }\href {\doibase 10.3847/2041-8213/ac089b}
  {\bibfield  {journal} {\bibinfo  {journal} {Astrophys. J. Lett.}\ }\textbf
  {\bibinfo {volume} {918}},\ \bibinfo {pages} {L28} (\bibinfo {year}
  {2021})},\ \Eprint {http://arxiv.org/abs/2105.06979} {arXiv:2105.06979
  [astro-ph.HE]} \BibitemShut {NoStop}%
\bibitem [{\citenamefont {Riley}\ \emph {et~al.}(2019)\citenamefont {Riley}
  \emph {et~al.}}]{Riley:2019yda}%
  \BibitemOpen
  \bibfield  {author} {\bibinfo {author} {\bibfnamefont {T.~E.}\ \bibnamefont
  {Riley}} \emph {et~al.},\ }\href {\doibase 10.3847/2041-8213/ab481c}
  {\bibfield  {journal} {\bibinfo  {journal} {Astrophys. J. Lett.}\ }\textbf
  {\bibinfo {volume} {887}},\ \bibinfo {pages} {L21} (\bibinfo {year}
  {2019})},\ \Eprint {http://arxiv.org/abs/1912.05702} {arXiv:1912.05702
  [astro-ph.HE]} \BibitemShut {NoStop}%
\bibitem [{\citenamefont {Riley}\ \emph {et~al.}(2021)\citenamefont {Riley}
  \emph {et~al.}}]{Riley:2021pdl}%
  \BibitemOpen
  \bibfield  {author} {\bibinfo {author} {\bibfnamefont {T.~E.}\ \bibnamefont
  {Riley}} \emph {et~al.},\ }\href {\doibase 10.3847/2041-8213/ac0a81}
  {\bibfield  {journal} {\bibinfo  {journal} {Astrophys. J. Lett.}\ }\textbf
  {\bibinfo {volume} {918}},\ \bibinfo {pages} {L27} (\bibinfo {year}
  {2021})},\ \Eprint {http://arxiv.org/abs/2105.06980} {arXiv:2105.06980
  [astro-ph.HE]} \BibitemShut {NoStop}%
\bibitem [{\citenamefont {Most}\ \emph {et~al.}(2019)\citenamefont {Most},
  \citenamefont {Papenfort}, \citenamefont {Dexheimer}, \citenamefont
  {Hanauske}, \citenamefont {Schramm}, \citenamefont {St\"ocker},\ and\
  \citenamefont {Rezzolla}}]{Most:2018eaw}%
  \BibitemOpen
  \bibfield  {author} {\bibinfo {author} {\bibfnamefont {E.~R.}\ \bibnamefont
  {Most}}, \bibinfo {author} {\bibfnamefont {L.~J.}\ \bibnamefont {Papenfort}},
  \bibinfo {author} {\bibfnamefont {V.}~\bibnamefont {Dexheimer}}, \bibinfo
  {author} {\bibfnamefont {M.}~\bibnamefont {Hanauske}}, \bibinfo {author}
  {\bibfnamefont {S.}~\bibnamefont {Schramm}}, \bibinfo {author} {\bibfnamefont
  {H.}~\bibnamefont {St\"ocker}}, \ and\ \bibinfo {author} {\bibfnamefont
  {L.}~\bibnamefont {Rezzolla}},\ }\href {\doibase
  10.1103/PhysRevLett.122.061101} {\bibfield  {journal} {\bibinfo  {journal}
  {Phys. Rev. Lett.}\ }\textbf {\bibinfo {volume} {122}},\ \bibinfo {pages}
  {061101} (\bibinfo {year} {2019})},\ \Eprint
  {http://arxiv.org/abs/1807.03684} {arXiv:1807.03684 [astro-ph.HE]}
  \BibitemShut {NoStop}%
\bibitem [{\citenamefont {Zha}\ \emph {et~al.}(2020)\citenamefont {Zha},
  \citenamefont {O'Connor}, \citenamefont {Chu}, \citenamefont {Lin},\ and\
  \citenamefont {Couch}}]{Zha:2020gjw}%
  \BibitemOpen
  \bibfield  {author} {\bibinfo {author} {\bibfnamefont {S.}~\bibnamefont
  {Zha}}, \bibinfo {author} {\bibfnamefont {E.~P.}\ \bibnamefont {O'Connor}},
  \bibinfo {author} {\bibfnamefont {M.-c.}\ \bibnamefont {Chu}}, \bibinfo
  {author} {\bibfnamefont {L.-M.}\ \bibnamefont {Lin}}, \ and\ \bibinfo
  {author} {\bibfnamefont {S.~M.}\ \bibnamefont {Couch}},\ }\href {\doibase
  10.1103/PhysRevLett.127.219901} {\bibfield  {journal} {\bibinfo  {journal}
  {Phys. Rev. Lett.}\ }\textbf {\bibinfo {volume} {125}},\ \bibinfo {pages}
  {051102} (\bibinfo {year} {2020})},\ \bibinfo {note} {[Erratum:
  Phys.Rev.Lett. 127, 219901 (2021)]},\ \Eprint
  {http://arxiv.org/abs/2007.04716} {arXiv:2007.04716 [astro-ph.HE]}
  \BibitemShut {NoStop}%
\bibitem [{\citenamefont {Borsanyi}\ \emph {et~al.}(2016)\citenamefont
  {Borsanyi} \emph {et~al.}}]{Borsanyi:2016ksw}%
  \BibitemOpen
  \bibfield  {author} {\bibinfo {author} {\bibfnamefont {S.}~\bibnamefont
  {Borsanyi}} \emph {et~al.},\ }\href {\doibase 10.1038/nature20115} {\bibfield
   {journal} {\bibinfo  {journal} {Nature}\ }\textbf {\bibinfo {volume}
  {539}},\ \bibinfo {pages} {69} (\bibinfo {year} {2016})},\ \Eprint
  {http://arxiv.org/abs/1606.07494} {arXiv:1606.07494 [hep-lat]} \BibitemShut
  {NoStop}%
\bibitem [{\citenamefont {Monnai}\ \emph {et~al.}(2019)\citenamefont {Monnai},
  \citenamefont {Schenke},\ and\ \citenamefont {Shen}}]{Monnai:2019hkn}%
  \BibitemOpen
  \bibfield  {author} {\bibinfo {author} {\bibfnamefont {A.}~\bibnamefont
  {Monnai}}, \bibinfo {author} {\bibfnamefont {B.}~\bibnamefont {Schenke}}, \
  and\ \bibinfo {author} {\bibfnamefont {C.}~\bibnamefont {Shen}},\ }\href
  {\doibase 10.1103/PhysRevC.100.024907} {\bibfield  {journal} {\bibinfo
  {journal} {Phys. Rev. C}\ }\textbf {\bibinfo {volume} {100}},\ \bibinfo
  {pages} {024907} (\bibinfo {year} {2019})},\ \Eprint
  {http://arxiv.org/abs/1902.05095} {arXiv:1902.05095 [nucl-th]} \BibitemShut
  {NoStop}%
\bibitem [{\citenamefont {Noronha-Hostler}\ \emph {et~al.}(2019)\citenamefont
  {Noronha-Hostler}, \citenamefont {Parotto}, \citenamefont {Ratti},\ and\
  \citenamefont {Stafford}}]{Noronha-Hostler:2019ayj}%
  \BibitemOpen
  \bibfield  {author} {\bibinfo {author} {\bibfnamefont {J.}~\bibnamefont
  {Noronha-Hostler}}, \bibinfo {author} {\bibfnamefont {P.}~\bibnamefont
  {Parotto}}, \bibinfo {author} {\bibfnamefont {C.}~\bibnamefont {Ratti}}, \
  and\ \bibinfo {author} {\bibfnamefont {J.~M.}\ \bibnamefont {Stafford}},\
  }\href {\doibase 10.1103/PhysRevC.100.064910} {\bibfield  {journal} {\bibinfo
   {journal} {Phys. Rev. C}\ }\textbf {\bibinfo {volume} {100}},\ \bibinfo
  {pages} {064910} (\bibinfo {year} {2019})},\ \Eprint
  {http://arxiv.org/abs/1902.06723} {arXiv:1902.06723 [hep-ph]} \BibitemShut
  {NoStop}%
\bibitem [{\citenamefont {Karthein}\ \emph {et~al.}(2021)\citenamefont
  {Karthein}, \citenamefont {Mroczek}, \citenamefont {Nava~Acuna},
  \citenamefont {Noronha-Hostler}, \citenamefont {Parotto}, \citenamefont
  {Price},\ and\ \citenamefont {Ratti}}]{Karthein:2021nxe}%
  \BibitemOpen
  \bibfield  {author} {\bibinfo {author} {\bibfnamefont {J.~M.}\ \bibnamefont
  {Karthein}}, \bibinfo {author} {\bibfnamefont {D.}~\bibnamefont {Mroczek}},
  \bibinfo {author} {\bibfnamefont {A.~R.}\ \bibnamefont {Nava~Acuna}},
  \bibinfo {author} {\bibfnamefont {J.}~\bibnamefont {Noronha-Hostler}},
  \bibinfo {author} {\bibfnamefont {P.}~\bibnamefont {Parotto}}, \bibinfo
  {author} {\bibfnamefont {D.~R.~P.}\ \bibnamefont {Price}}, \ and\ \bibinfo
  {author} {\bibfnamefont {C.}~\bibnamefont {Ratti}},\ }\href {\doibase
  10.1140/epjp/s13360-021-01615-5} {\bibfield  {journal} {\bibinfo  {journal}
  {Eur. Phys. J. Plus}\ }\textbf {\bibinfo {volume} {136}},\ \bibinfo {pages}
  {621} (\bibinfo {year} {2021})},\ \Eprint {http://arxiv.org/abs/2103.08146}
  {arXiv:2103.08146 [hep-ph]} \BibitemShut {NoStop}%
\bibitem [{\citenamefont {Xu}(2017)}]{Xu:2016hxf}%
  \BibitemOpen
  \bibfield  {author} {\bibinfo {author} {\bibfnamefont {J.}~\bibnamefont {Xu}}
  (\bibinfo {collaboration} {STAR}),\ }\href {\doibase
  10.1088/1742-6596/779/1/012073} {\bibfield  {journal} {\bibinfo  {journal}
  {J. Phys. Conf. Ser.}\ }\textbf {\bibinfo {volume} {779}},\ \bibinfo {pages}
  {012073} (\bibinfo {year} {2017})},\ \Eprint
  {http://arxiv.org/abs/1611.07132} {arXiv:1611.07132 [hep-ex]} \BibitemShut
  {NoStop}%
\bibitem [{\citenamefont {Noronha-Hostler}\ \emph {et~al.}(2016)\citenamefont
  {Noronha-Hostler}, \citenamefont {Bellwied}, \citenamefont {Gunther},
  \citenamefont {Parotto}, \citenamefont {Pasztor}, \citenamefont
  {Portillo~Vazquez},\ and\ \citenamefont {Ratti}}]{Noronha-Hostler:2016rpd}%
  \BibitemOpen
  \bibfield  {author} {\bibinfo {author} {\bibfnamefont {J.}~\bibnamefont
  {Noronha-Hostler}}, \bibinfo {author} {\bibfnamefont {R.}~\bibnamefont
  {Bellwied}}, \bibinfo {author} {\bibfnamefont {J.}~\bibnamefont {Gunther}},
  \bibinfo {author} {\bibfnamefont {P.}~\bibnamefont {Parotto}}, \bibinfo
  {author} {\bibfnamefont {A.}~\bibnamefont {Pasztor}}, \bibinfo {author}
  {\bibfnamefont {I.}~\bibnamefont {Portillo~Vazquez}}, \ and\ \bibinfo
  {author} {\bibfnamefont {C.}~\bibnamefont {Ratti}},\ }\href@noop {} {\
  (\bibinfo {year} {2016})},\ \Eprint {http://arxiv.org/abs/1607.02527}
  {arXiv:1607.02527 [hep-ph]} \BibitemShut {NoStop}%
\bibitem [{\citenamefont {Bluhm}\ and\ \citenamefont
  {Nahrgang}(2019)}]{Bluhm:2018aei}%
  \BibitemOpen
  \bibfield  {author} {\bibinfo {author} {\bibfnamefont {M.}~\bibnamefont
  {Bluhm}}\ and\ \bibinfo {author} {\bibfnamefont {M.}~\bibnamefont
  {Nahrgang}},\ }\href {\doibase 10.1140/epjc/s10052-019-6661-3} {\bibfield
  {journal} {\bibinfo  {journal} {Eur. Phys. J. C}\ }\textbf {\bibinfo {volume}
  {79}},\ \bibinfo {pages} {155} (\bibinfo {year} {2019})},\ \Eprint
  {http://arxiv.org/abs/1806.04499} {arXiv:1806.04499 [nucl-th]} \BibitemShut
  {NoStop}%
\bibitem [{\citenamefont {Alba}\ \emph {et~al.}(2017)\citenamefont {Alba} \emph
  {et~al.}}]{Alba:2017mqu}%
  \BibitemOpen
  \bibfield  {author} {\bibinfo {author} {\bibfnamefont {P.}~\bibnamefont
  {Alba}} \emph {et~al.},\ }\href {\doibase 10.1103/PhysRevD.96.034517}
  {\bibfield  {journal} {\bibinfo  {journal} {Phys. Rev. D}\ }\textbf {\bibinfo
  {volume} {96}},\ \bibinfo {pages} {034517} (\bibinfo {year} {2017})},\
  \Eprint {http://arxiv.org/abs/1702.01113} {arXiv:1702.01113 [hep-lat]}
  \BibitemShut {NoStop}%
\bibitem [{\citenamefont {Bellwied}\ \emph {et~al.}(2019)\citenamefont
  {Bellwied}, \citenamefont {Noronha-Hostler}, \citenamefont {Parotto},
  \citenamefont {Portillo~Vazquez}, \citenamefont {Ratti},\ and\ \citenamefont
  {Stafford}}]{Bellwied:2018tkc}%
  \BibitemOpen
  \bibfield  {author} {\bibinfo {author} {\bibfnamefont {R.}~\bibnamefont
  {Bellwied}}, \bibinfo {author} {\bibfnamefont {J.}~\bibnamefont
  {Noronha-Hostler}}, \bibinfo {author} {\bibfnamefont {P.}~\bibnamefont
  {Parotto}}, \bibinfo {author} {\bibfnamefont {I.}~\bibnamefont
  {Portillo~Vazquez}}, \bibinfo {author} {\bibfnamefont {C.}~\bibnamefont
  {Ratti}}, \ and\ \bibinfo {author} {\bibfnamefont {J.~M.}\ \bibnamefont
  {Stafford}},\ }\href {\doibase 10.1103/PhysRevC.99.034912} {\bibfield
  {journal} {\bibinfo  {journal} {Phys. Rev. C}\ }\textbf {\bibinfo {volume}
  {99}},\ \bibinfo {pages} {034912} (\bibinfo {year} {2019})},\ \Eprint
  {http://arxiv.org/abs/1805.00088} {arXiv:1805.00088 [hep-ph]} \BibitemShut
  {NoStop}%
\bibitem [{\citenamefont {Bellwied}\ \emph {et~al.}(2020)\citenamefont
  {Bellwied}, \citenamefont {Borsanyi}, \citenamefont {Fodor}, \citenamefont
  {Guenther}, \citenamefont {Noronha-Hostler}, \citenamefont {Parotto},
  \citenamefont {Pasztor}, \citenamefont {Ratti},\ and\ \citenamefont
  {Stafford}}]{Bellwied:2019pxh}%
  \BibitemOpen
  \bibfield  {author} {\bibinfo {author} {\bibfnamefont {R.}~\bibnamefont
  {Bellwied}}, \bibinfo {author} {\bibfnamefont {S.}~\bibnamefont {Borsanyi}},
  \bibinfo {author} {\bibfnamefont {Z.}~\bibnamefont {Fodor}}, \bibinfo
  {author} {\bibfnamefont {J.~N.}\ \bibnamefont {Guenther}}, \bibinfo {author}
  {\bibfnamefont {J.}~\bibnamefont {Noronha-Hostler}}, \bibinfo {author}
  {\bibfnamefont {P.}~\bibnamefont {Parotto}}, \bibinfo {author} {\bibfnamefont
  {A.}~\bibnamefont {Pasztor}}, \bibinfo {author} {\bibfnamefont
  {C.}~\bibnamefont {Ratti}}, \ and\ \bibinfo {author} {\bibfnamefont {J.~M.}\
  \bibnamefont {Stafford}},\ }\href {\doibase 10.1103/PhysRevD.101.034506}
  {\bibfield  {journal} {\bibinfo  {journal} {Phys. Rev. D}\ }\textbf {\bibinfo
  {volume} {101}},\ \bibinfo {pages} {034506} (\bibinfo {year} {2020})},\
  \Eprint {http://arxiv.org/abs/1910.14592} {arXiv:1910.14592 [hep-lat]}
  \BibitemShut {NoStop}%
\bibitem [{\citenamefont {Alba}\ \emph {et~al.}(2020)\citenamefont {Alba},
  \citenamefont {Sarti}, \citenamefont {Noronha-Hostler}, \citenamefont
  {Parotto}, \citenamefont {Portillo-Vazquez}, \citenamefont {Ratti},\ and\
  \citenamefont {Stafford}}]{Alba:2020jir}%
  \BibitemOpen
  \bibfield  {author} {\bibinfo {author} {\bibfnamefont {P.}~\bibnamefont
  {Alba}}, \bibinfo {author} {\bibfnamefont {V.~M.}\ \bibnamefont {Sarti}},
  \bibinfo {author} {\bibfnamefont {J.}~\bibnamefont {Noronha-Hostler}},
  \bibinfo {author} {\bibfnamefont {P.}~\bibnamefont {Parotto}}, \bibinfo
  {author} {\bibfnamefont {I.}~\bibnamefont {Portillo-Vazquez}}, \bibinfo
  {author} {\bibfnamefont {C.}~\bibnamefont {Ratti}}, \ and\ \bibinfo {author}
  {\bibfnamefont {J.~M.}\ \bibnamefont {Stafford}},\ }\href {\doibase
  10.1103/PhysRevC.101.054905} {\bibfield  {journal} {\bibinfo  {journal}
  {Phys. Rev. C}\ }\textbf {\bibinfo {volume} {101}},\ \bibinfo {pages}
  {054905} (\bibinfo {year} {2020})},\ \Eprint
  {http://arxiv.org/abs/2002.12395} {arXiv:2002.12395 [hep-ph]} \BibitemShut
  {NoStop}%
\bibitem [{\citenamefont {San~Martin}\ \emph {et~al.}(2023)\citenamefont
  {San~Martin}, \citenamefont {Hirayama}, \citenamefont {Hammelmann},
  \citenamefont {Karthein}, \citenamefont {Parotto}, \citenamefont
  {Noronha-Hostler}, \citenamefont {Ratti},\ and\ \citenamefont
  {Elfner}}]{SanMartin:2023zhv}%
  \BibitemOpen
  \bibfield  {author} {\bibinfo {author} {\bibfnamefont {J.~S.}\ \bibnamefont
  {San~Martin}}, \bibinfo {author} {\bibfnamefont {R.}~\bibnamefont
  {Hirayama}}, \bibinfo {author} {\bibfnamefont {J.}~\bibnamefont
  {Hammelmann}}, \bibinfo {author} {\bibfnamefont {J.~M.}\ \bibnamefont
  {Karthein}}, \bibinfo {author} {\bibfnamefont {P.}~\bibnamefont {Parotto}},
  \bibinfo {author} {\bibfnamefont {J.}~\bibnamefont {Noronha-Hostler}},
  \bibinfo {author} {\bibfnamefont {C.}~\bibnamefont {Ratti}}, \ and\ \bibinfo
  {author} {\bibfnamefont {H.}~\bibnamefont {Elfner}},\ }\href@noop {} {\
  (\bibinfo {year} {2023})},\ \Eprint {http://arxiv.org/abs/2309.01737}
  {arXiv:2309.01737 [nucl-th]} \BibitemShut {NoStop}%
\bibitem [{\citenamefont {Kaplan}\ and\ \citenamefont
  {Nelson}(1986)}]{Kaplan:1986yq}%
  \BibitemOpen
  \bibfield  {author} {\bibinfo {author} {\bibfnamefont {D.~B.}\ \bibnamefont
  {Kaplan}}\ and\ \bibinfo {author} {\bibfnamefont {A.~E.}\ \bibnamefont
  {Nelson}},\ }\href {\doibase 10.1016/0370-2693(86)90331-X} {\bibfield
  {journal} {\bibinfo  {journal} {Phys. Lett. B}\ }\textbf {\bibinfo {volume}
  {175}},\ \bibinfo {pages} {57} (\bibinfo {year} {1986})}\BibitemShut
  {NoStop}%
\bibitem [{\citenamefont {Glendenning}(1992)}]{Glendenning:1992vb}%
  \BibitemOpen
  \bibfield  {author} {\bibinfo {author} {\bibfnamefont {N.~K.}\ \bibnamefont
  {Glendenning}},\ }\href {\doibase 10.1103/PhysRevD.46.1274} {\bibfield
  {journal} {\bibinfo  {journal} {Phys. Rev. D}\ }\textbf {\bibinfo {volume}
  {46}},\ \bibinfo {pages} {1274} (\bibinfo {year} {1992})}\BibitemShut
  {NoStop}%
\bibitem [{\citenamefont {Schaffner}\ \emph {et~al.}(1994)\citenamefont
  {Schaffner}, \citenamefont {Dover}, \citenamefont {Gal}, \citenamefont
  {Greiner}, \citenamefont {Millener},\ and\ \citenamefont
  {Stoecker}}]{Schaffner:1993qj}%
  \BibitemOpen
  \bibfield  {author} {\bibinfo {author} {\bibfnamefont {J.}~\bibnamefont
  {Schaffner}}, \bibinfo {author} {\bibfnamefont {C.~B.}\ \bibnamefont
  {Dover}}, \bibinfo {author} {\bibfnamefont {A.}~\bibnamefont {Gal}}, \bibinfo
  {author} {\bibfnamefont {C.}~\bibnamefont {Greiner}}, \bibinfo {author}
  {\bibfnamefont {D.~J.}\ \bibnamefont {Millener}}, \ and\ \bibinfo {author}
  {\bibfnamefont {H.}~\bibnamefont {Stoecker}},\ }\href {\doibase
  10.1006/aphy.1994.1090} {\bibfield  {journal} {\bibinfo  {journal} {Annals
  Phys.}\ }\textbf {\bibinfo {volume} {235}},\ \bibinfo {pages} {35} (\bibinfo
  {year} {1994})}\BibitemShut {NoStop}%
\bibitem [{\citenamefont {Schaffner}\ \emph {et~al.}(1993)\citenamefont
  {Schaffner}, \citenamefont {Dover}, \citenamefont {Gal}, \citenamefont
  {Greiner},\ and\ \citenamefont {Stoecker}}]{Schaffner:1993nn}%
  \BibitemOpen
  \bibfield  {author} {\bibinfo {author} {\bibfnamefont {J.}~\bibnamefont
  {Schaffner}}, \bibinfo {author} {\bibfnamefont {C.~B.}\ \bibnamefont
  {Dover}}, \bibinfo {author} {\bibfnamefont {A.}~\bibnamefont {Gal}}, \bibinfo
  {author} {\bibfnamefont {C.}~\bibnamefont {Greiner}}, \ and\ \bibinfo
  {author} {\bibfnamefont {H.}~\bibnamefont {Stoecker}},\ }\href {\doibase
  10.1103/PhysRevLett.71.1328} {\bibfield  {journal} {\bibinfo  {journal}
  {Phys. Rev. Lett.}\ }\textbf {\bibinfo {volume} {71}},\ \bibinfo {pages}
  {1328} (\bibinfo {year} {1993})}\BibitemShut {NoStop}%
\bibitem [{\citenamefont {Hatsuda}\ and\ \citenamefont
  {Kunihiro}(1994)}]{Hatsuda:1994pi}%
  \BibitemOpen
  \bibfield  {author} {\bibinfo {author} {\bibfnamefont {T.}~\bibnamefont
  {Hatsuda}}\ and\ \bibinfo {author} {\bibfnamefont {T.}~\bibnamefont
  {Kunihiro}},\ }\href {\doibase 10.1016/0370-1573(94)90022-1} {\bibfield
  {journal} {\bibinfo  {journal} {Phys. Rept.}\ }\textbf {\bibinfo {volume}
  {247}},\ \bibinfo {pages} {221} (\bibinfo {year} {1994})},\ \Eprint
  {http://arxiv.org/abs/hep-ph/9401310} {arXiv:hep-ph/9401310} \BibitemShut
  {NoStop}%
\bibitem [{\citenamefont {Pisarski}(2000)}]{Pisarski:1999gq}%
  \BibitemOpen
  \bibfield  {author} {\bibinfo {author} {\bibfnamefont {R.~D.}\ \bibnamefont
  {Pisarski}},\ }\href {\doibase 10.1103/PhysRevC.62.035202} {\bibfield
  {journal} {\bibinfo  {journal} {Phys. Rev. C}\ }\textbf {\bibinfo {volume}
  {62}},\ \bibinfo {pages} {035202} (\bibinfo {year} {2000})},\ \Eprint
  {http://arxiv.org/abs/nucl-th/9912070} {arXiv:nucl-th/9912070} \BibitemShut
  {NoStop}%
\bibitem [{\citenamefont {Buballa}(2005)}]{Buballa:2003qv}%
  \BibitemOpen
  \bibfield  {author} {\bibinfo {author} {\bibfnamefont {M.}~\bibnamefont
  {Buballa}},\ }\href {\doibase 10.1016/j.physrep.2004.11.004} {\bibfield
  {journal} {\bibinfo  {journal} {Phys. Rept.}\ }\textbf {\bibinfo {volume}
  {407}},\ \bibinfo {pages} {205} (\bibinfo {year} {2005})},\ \Eprint
  {http://arxiv.org/abs/hep-ph/0402234} {arXiv:hep-ph/0402234} \BibitemShut
  {NoStop}%
\bibitem [{\citenamefont {Weber}(2005)}]{Weber:2004kj}%
  \BibitemOpen
  \bibfield  {author} {\bibinfo {author} {\bibfnamefont {F.}~\bibnamefont
  {Weber}},\ }\href {\doibase 10.1016/j.ppnp.2004.07.001} {\bibfield  {journal}
  {\bibinfo  {journal} {Prog. Part. Nucl. Phys.}\ }\textbf {\bibinfo {volume}
  {54}},\ \bibinfo {pages} {193} (\bibinfo {year} {2005})},\ \Eprint
  {http://arxiv.org/abs/astro-ph/0407155} {arXiv:astro-ph/0407155} \BibitemShut
  {NoStop}%
\bibitem [{\citenamefont {Weissenborn}\ \emph {et~al.}(2012)\citenamefont
  {Weissenborn}, \citenamefont {Chatterjee},\ and\ \citenamefont
  {Schaffner-Bielich}}]{Weissenborn:2011kb}%
  \BibitemOpen
  \bibfield  {author} {\bibinfo {author} {\bibfnamefont {S.}~\bibnamefont
  {Weissenborn}}, \bibinfo {author} {\bibfnamefont {D.}~\bibnamefont
  {Chatterjee}}, \ and\ \bibinfo {author} {\bibfnamefont {J.}~\bibnamefont
  {Schaffner-Bielich}},\ }\href {\doibase 10.1016/j.nuclphysa.2012.02.012}
  {\bibfield  {journal} {\bibinfo  {journal} {Nucl. Phys. A}\ }\textbf
  {\bibinfo {volume} {881}},\ \bibinfo {pages} {62} (\bibinfo {year} {2012})},\
  \Eprint {http://arxiv.org/abs/1111.6049} {arXiv:1111.6049 [astro-ph.HE]}
  \BibitemShut {NoStop}%
\bibitem [{\citenamefont {Dexheimer}\ \emph {et~al.}(2015)\citenamefont
  {Dexheimer}, \citenamefont {Negreiros},\ and\ \citenamefont
  {Schramm}}]{Dexheimer:2014pea}%
  \BibitemOpen
  \bibfield  {author} {\bibinfo {author} {\bibfnamefont {V.}~\bibnamefont
  {Dexheimer}}, \bibinfo {author} {\bibfnamefont {R.}~\bibnamefont
  {Negreiros}}, \ and\ \bibinfo {author} {\bibfnamefont {S.}~\bibnamefont
  {Schramm}},\ }\href {\doibase 10.1103/PhysRevC.91.055808} {\bibfield
  {journal} {\bibinfo  {journal} {Phys. Rev. C}\ }\textbf {\bibinfo {volume}
  {91}},\ \bibinfo {pages} {055808} (\bibinfo {year} {2015})},\ \Eprint
  {http://arxiv.org/abs/1411.4623} {arXiv:1411.4623 [astro-ph.HE]} \BibitemShut
  {NoStop}%
\bibitem [{\citenamefont {Gomes}\ \emph {et~al.}(2015)\citenamefont {Gomes},
  \citenamefont {Dexheimer}, \citenamefont {Schramm},\ and\ \citenamefont
  {Vasconcellos}}]{Gomes:2014aka}%
  \BibitemOpen
  \bibfield  {author} {\bibinfo {author} {\bibfnamefont {R.~O.}\ \bibnamefont
  {Gomes}}, \bibinfo {author} {\bibfnamefont {V.}~\bibnamefont {Dexheimer}},
  \bibinfo {author} {\bibfnamefont {S.}~\bibnamefont {Schramm}}, \ and\
  \bibinfo {author} {\bibfnamefont {C.~A.~Z.}\ \bibnamefont {Vasconcellos}},\
  }\href {\doibase 10.1088/0004-637X/808/1/8} {\bibfield  {journal} {\bibinfo
  {journal} {Astrophys. J.}\ }\textbf {\bibinfo {volume} {808}},\ \bibinfo
  {pages} {8} (\bibinfo {year} {2015})},\ \Eprint
  {http://arxiv.org/abs/1411.4875} {arXiv:1411.4875 [astro-ph.SR]} \BibitemShut
  {NoStop}%
\bibitem [{\citenamefont {Tolos}\ and\ \citenamefont
  {Fabbietti}(2020)}]{Tolos:2020aln}%
  \BibitemOpen
  \bibfield  {author} {\bibinfo {author} {\bibfnamefont {L.}~\bibnamefont
  {Tolos}}\ and\ \bibinfo {author} {\bibfnamefont {L.}~\bibnamefont
  {Fabbietti}},\ }\href {\doibase 10.1016/j.ppnp.2020.103770} {\bibfield
  {journal} {\bibinfo  {journal} {Prog. Part. Nucl. Phys.}\ }\textbf {\bibinfo
  {volume} {112}},\ \bibinfo {pages} {103770} (\bibinfo {year} {2020})},\
  \Eprint {http://arxiv.org/abs/2002.09223} {arXiv:2002.09223 [nucl-ex]}
  \BibitemShut {NoStop}%
\bibitem [{\citenamefont {Mroczek}\ \emph {et~al.}(2024)\citenamefont
  {Mroczek}, \citenamefont {Miller}, \citenamefont {Noronha-Hostler},\ and\
  \citenamefont {Yunes}}]{Mroczek:2023zxo}%
  \BibitemOpen
  \bibfield  {author} {\bibinfo {author} {\bibfnamefont {D.}~\bibnamefont
  {Mroczek}}, \bibinfo {author} {\bibfnamefont {M.~C.}\ \bibnamefont {Miller}},
  \bibinfo {author} {\bibfnamefont {J.}~\bibnamefont {Noronha-Hostler}}, \ and\
  \bibinfo {author} {\bibfnamefont {N.}~\bibnamefont {Yunes}},\ }\href
  {\doibase 10.1103/PhysRevD.110.123009} {\bibfield  {journal} {\bibinfo
  {journal} {Phys. Rev. D}\ }\textbf {\bibinfo {volume} {110}},\ \bibinfo
  {pages} {123009} (\bibinfo {year} {2024})},\ \Eprint
  {http://arxiv.org/abs/2309.02345} {arXiv:2309.02345 [astro-ph.HE]}
  \BibitemShut {NoStop}%
\bibitem [{\citenamefont {Wiringa}\ \emph {et~al.}(1988)\citenamefont
  {Wiringa}, \citenamefont {Fiks},\ and\ \citenamefont
  {Fabrocini}}]{Wiringa:1988tp}%
  \BibitemOpen
  \bibfield  {author} {\bibinfo {author} {\bibfnamefont {R.~B.}\ \bibnamefont
  {Wiringa}}, \bibinfo {author} {\bibfnamefont {V.}~\bibnamefont {Fiks}}, \
  and\ \bibinfo {author} {\bibfnamefont {A.}~\bibnamefont {Fabrocini}},\ }\href
  {\doibase 10.1103/PhysRevC.38.1010} {\bibfield  {journal} {\bibinfo
  {journal} {Phys. Rev. C}\ }\textbf {\bibinfo {volume} {38}},\ \bibinfo
  {pages} {1010} (\bibinfo {year} {1988})}\BibitemShut {NoStop}%
\bibitem [{\citenamefont {Steiner}(2006)}]{Steiner:2006bx}%
  \BibitemOpen
  \bibfield  {author} {\bibinfo {author} {\bibfnamefont {A.~W.}\ \bibnamefont
  {Steiner}},\ }\href {\doibase 10.1103/PhysRevC.74.045808} {\bibfield
  {journal} {\bibinfo  {journal} {Phys. Rev. C}\ }\textbf {\bibinfo {volume}
  {74}},\ \bibinfo {pages} {045808} (\bibinfo {year} {2006})},\ \Eprint
  {http://arxiv.org/abs/nucl-th/0607040} {arXiv:nucl-th/0607040} \BibitemShut
  {NoStop}%
\bibitem [{\citenamefont {Chen}\ \emph {et~al.}(2007)\citenamefont {Chen},
  \citenamefont {Ko},\ and\ \citenamefont {Li}}]{Chen:2007ih}%
  \BibitemOpen
  \bibfield  {author} {\bibinfo {author} {\bibfnamefont {L.-W.}\ \bibnamefont
  {Chen}}, \bibinfo {author} {\bibfnamefont {C.~M.}\ \bibnamefont {Ko}}, \ and\
  \bibinfo {author} {\bibfnamefont {B.-A.}\ \bibnamefont {Li}},\ }\href
  {\doibase 10.1103/PhysRevC.76.054316} {\bibfield  {journal} {\bibinfo
  {journal} {Phys. Rev. C}\ }\textbf {\bibinfo {volume} {76}},\ \bibinfo
  {pages} {054316} (\bibinfo {year} {2007})},\ \Eprint
  {http://arxiv.org/abs/0709.0900} {arXiv:0709.0900 [nucl-th]} \BibitemShut
  {NoStop}%
\bibitem [{\citenamefont {Vidana}\ \emph {et~al.}(2009)\citenamefont {Vidana},
  \citenamefont {Providencia}, \citenamefont {Polls},\ and\ \citenamefont
  {Rios}}]{Vidana:2009is}%
  \BibitemOpen
  \bibfield  {author} {\bibinfo {author} {\bibfnamefont {I.}~\bibnamefont
  {Vidana}}, \bibinfo {author} {\bibfnamefont {C.}~\bibnamefont {Providencia}},
  \bibinfo {author} {\bibfnamefont {A.}~\bibnamefont {Polls}}, \ and\ \bibinfo
  {author} {\bibfnamefont {A.}~\bibnamefont {Rios}},\ }\href {\doibase
  10.1103/PhysRevC.80.045806} {\bibfield  {journal} {\bibinfo  {journal} {Phys.
  Rev. C}\ }\textbf {\bibinfo {volume} {80}},\ \bibinfo {pages} {045806}
  (\bibinfo {year} {2009})},\ \Eprint {http://arxiv.org/abs/0907.1165}
  {arXiv:0907.1165 [nucl-th]} \BibitemShut {NoStop}%
\bibitem [{\citenamefont {Hebeler}\ \emph {et~al.}(2010)\citenamefont
  {Hebeler}, \citenamefont {Lattimer}, \citenamefont {Pethick},\ and\
  \citenamefont {Schwenk}}]{Hebeler:2010jx}%
  \BibitemOpen
  \bibfield  {author} {\bibinfo {author} {\bibfnamefont {K.}~\bibnamefont
  {Hebeler}}, \bibinfo {author} {\bibfnamefont {J.~M.}\ \bibnamefont
  {Lattimer}}, \bibinfo {author} {\bibfnamefont {C.~J.}\ \bibnamefont
  {Pethick}}, \ and\ \bibinfo {author} {\bibfnamefont {A.}~\bibnamefont
  {Schwenk}},\ }\href {\doibase 10.1103/PhysRevLett.105.161102} {\bibfield
  {journal} {\bibinfo  {journal} {Phys. Rev. Lett.}\ }\textbf {\bibinfo
  {volume} {105}},\ \bibinfo {pages} {161102} (\bibinfo {year} {2010})},\
  \Eprint {http://arxiv.org/abs/1007.1746} {arXiv:1007.1746 [nucl-th]}
  \BibitemShut {NoStop}%
\bibitem [{\citenamefont {Cai}\ and\ \citenamefont {Chen}(2012)}]{Cai:2011zn}%
  \BibitemOpen
  \bibfield  {author} {\bibinfo {author} {\bibfnamefont {B.-J.}\ \bibnamefont
  {Cai}}\ and\ \bibinfo {author} {\bibfnamefont {L.-W.}\ \bibnamefont {Chen}},\
  }\href {\doibase 10.1103/PhysRevC.85.024302} {\bibfield  {journal} {\bibinfo
  {journal} {Phys. Rev. C}\ }\textbf {\bibinfo {volume} {85}},\ \bibinfo
  {pages} {024302} (\bibinfo {year} {2012})},\ \Eprint
  {http://arxiv.org/abs/1111.4124} {arXiv:1111.4124 [nucl-th]} \BibitemShut
  {NoStop}%
\bibitem [{\citenamefont {Drischler}\ \emph {et~al.}(2014)\citenamefont
  {Drischler}, \citenamefont {Soma},\ and\ \citenamefont
  {Schwenk}}]{Drischler:2013iza}%
  \BibitemOpen
  \bibfield  {author} {\bibinfo {author} {\bibfnamefont {C.}~\bibnamefont
  {Drischler}}, \bibinfo {author} {\bibfnamefont {V.}~\bibnamefont {Soma}}, \
  and\ \bibinfo {author} {\bibfnamefont {A.}~\bibnamefont {Schwenk}},\ }\href
  {\doibase 10.1103/PhysRevC.89.025806} {\bibfield  {journal} {\bibinfo
  {journal} {Phys. Rev. C}\ }\textbf {\bibinfo {volume} {89}},\ \bibinfo
  {pages} {025806} (\bibinfo {year} {2014})},\ \Eprint
  {http://arxiv.org/abs/1310.5627} {arXiv:1310.5627 [nucl-th]} \BibitemShut
  {NoStop}%
\bibitem [{\citenamefont {Seif}\ and\ \citenamefont
  {Basu}(2014)}]{Seif:2013tja}%
  \BibitemOpen
  \bibfield  {author} {\bibinfo {author} {\bibfnamefont {W.~M.}\ \bibnamefont
  {Seif}}\ and\ \bibinfo {author} {\bibfnamefont {D.~N.}\ \bibnamefont
  {Basu}},\ }\href {\doibase 10.1103/PhysRevC.89.028801} {\bibfield  {journal}
  {\bibinfo  {journal} {Phys. Rev. C}\ }\textbf {\bibinfo {volume} {89}},\
  \bibinfo {pages} {028801} (\bibinfo {year} {2014})},\ \Eprint
  {http://arxiv.org/abs/1401.0090} {arXiv:1401.0090 [nucl-th]} \BibitemShut
  {NoStop}%
\bibitem [{\citenamefont {Gandolfi}\ \emph {et~al.}(2014)\citenamefont
  {Gandolfi}, \citenamefont {Carlson}, \citenamefont {Reddy}, \citenamefont
  {Steiner},\ and\ \citenamefont {Wiringa}}]{Gandolfi:2013baa}%
  \BibitemOpen
  \bibfield  {author} {\bibinfo {author} {\bibfnamefont {S.}~\bibnamefont
  {Gandolfi}}, \bibinfo {author} {\bibfnamefont {J.}~\bibnamefont {Carlson}},
  \bibinfo {author} {\bibfnamefont {S.}~\bibnamefont {Reddy}}, \bibinfo
  {author} {\bibfnamefont {A.~W.}\ \bibnamefont {Steiner}}, \ and\ \bibinfo
  {author} {\bibfnamefont {R.~B.}\ \bibnamefont {Wiringa}},\ }\href {\doibase
  10.1140/epja/i2014-14010-5} {\bibfield  {journal} {\bibinfo  {journal} {Eur.
  Phys. J. A}\ }\textbf {\bibinfo {volume} {50}},\ \bibinfo {pages} {10}
  (\bibinfo {year} {2014})},\ \Eprint {http://arxiv.org/abs/1307.5815}
  {arXiv:1307.5815 [nucl-th]} \BibitemShut {NoStop}%
\bibitem [{\citenamefont {Drischler}\ \emph {et~al.}(2016)\citenamefont
  {Drischler}, \citenamefont {Hebeler},\ and\ \citenamefont
  {Schwenk}}]{Drischler:2015eba}%
  \BibitemOpen
  \bibfield  {author} {\bibinfo {author} {\bibfnamefont {C.}~\bibnamefont
  {Drischler}}, \bibinfo {author} {\bibfnamefont {K.}~\bibnamefont {Hebeler}},
  \ and\ \bibinfo {author} {\bibfnamefont {A.}~\bibnamefont {Schwenk}},\ }\href
  {\doibase 10.1103/PhysRevC.93.054314} {\bibfield  {journal} {\bibinfo
  {journal} {Phys. Rev. C}\ }\textbf {\bibinfo {volume} {93}},\ \bibinfo
  {pages} {054314} (\bibinfo {year} {2016})},\ \Eprint
  {http://arxiv.org/abs/1510.06728} {arXiv:1510.06728 [nucl-th]} \BibitemShut
  {NoStop}%
\bibitem [{\citenamefont {Wellenhofer}\ \emph {et~al.}(2015)\citenamefont
  {Wellenhofer}, \citenamefont {Holt},\ and\ \citenamefont
  {Kaiser}}]{Wellenhofer:2015qba}%
  \BibitemOpen
  \bibfield  {author} {\bibinfo {author} {\bibfnamefont {C.}~\bibnamefont
  {Wellenhofer}}, \bibinfo {author} {\bibfnamefont {J.~W.}\ \bibnamefont
  {Holt}}, \ and\ \bibinfo {author} {\bibfnamefont {N.}~\bibnamefont
  {Kaiser}},\ }\href {\doibase 10.1103/PhysRevC.92.015801} {\bibfield
  {journal} {\bibinfo  {journal} {Phys. Rev. C}\ }\textbf {\bibinfo {volume}
  {92}},\ \bibinfo {pages} {015801} (\bibinfo {year} {2015})},\ \Eprint
  {http://arxiv.org/abs/1504.00177} {arXiv:1504.00177 [nucl-th]} \BibitemShut
  {NoStop}%
\bibitem [{\citenamefont {Wellenhofer}\ \emph {et~al.}(2016)\citenamefont
  {Wellenhofer}, \citenamefont {Holt},\ and\ \citenamefont
  {Kaiser}}]{Wellenhofer:2016lnl}%
  \BibitemOpen
  \bibfield  {author} {\bibinfo {author} {\bibfnamefont {C.}~\bibnamefont
  {Wellenhofer}}, \bibinfo {author} {\bibfnamefont {J.~W.}\ \bibnamefont
  {Holt}}, \ and\ \bibinfo {author} {\bibfnamefont {N.}~\bibnamefont
  {Kaiser}},\ }\href {\doibase 10.1103/PhysRevC.93.055802} {\bibfield
  {journal} {\bibinfo  {journal} {Phys. Rev. C}\ }\textbf {\bibinfo {volume}
  {93}},\ \bibinfo {pages} {055802} (\bibinfo {year} {2016})},\ \Eprint
  {http://arxiv.org/abs/1603.02935} {arXiv:1603.02935 [nucl-th]} \BibitemShut
  {NoStop}%
\bibitem [{\citenamefont {Drischler}\ \emph {et~al.}(2019)\citenamefont
  {Drischler}, \citenamefont {Hebeler},\ and\ \citenamefont
  {Schwenk}}]{Drischler:2017wtt}%
  \BibitemOpen
  \bibfield  {author} {\bibinfo {author} {\bibfnamefont {C.}~\bibnamefont
  {Drischler}}, \bibinfo {author} {\bibfnamefont {K.}~\bibnamefont {Hebeler}},
  \ and\ \bibinfo {author} {\bibfnamefont {A.}~\bibnamefont {Schwenk}},\ }\href
  {\doibase 10.1103/PhysRevLett.122.042501} {\bibfield  {journal} {\bibinfo
  {journal} {Phys. Rev. Lett.}\ }\textbf {\bibinfo {volume} {122}},\ \bibinfo
  {pages} {042501} (\bibinfo {year} {2019})},\ \Eprint
  {http://arxiv.org/abs/1710.08220} {arXiv:1710.08220 [nucl-th]} \BibitemShut
  {NoStop}%
\bibitem [{\citenamefont {Zhang}\ \emph {et~al.}(2018)\citenamefont {Zhang},
  \citenamefont {Li},\ and\ \citenamefont {Xu}}]{Zhang:2018vrx}%
  \BibitemOpen
  \bibfield  {author} {\bibinfo {author} {\bibfnamefont {N.-B.}\ \bibnamefont
  {Zhang}}, \bibinfo {author} {\bibfnamefont {B.-A.}\ \bibnamefont {Li}}, \
  and\ \bibinfo {author} {\bibfnamefont {J.}~\bibnamefont {Xu}},\ }\href
  {\doibase 10.3847/1538-4357/aac027} {\bibfield  {journal} {\bibinfo
  {journal} {Astrophys. J.}\ }\textbf {\bibinfo {volume} {859}},\ \bibinfo
  {pages} {90} (\bibinfo {year} {2018})},\ \Eprint
  {http://arxiv.org/abs/1801.06855} {arXiv:1801.06855 [nucl-th]} \BibitemShut
  {NoStop}%
\bibitem [{\citenamefont {Li}\ \emph {et~al.}(2019)\citenamefont {Li},
  \citenamefont {Krastev}, \citenamefont {Wen},\ and\ \citenamefont
  {Zhang}}]{Li:2019xxz}%
  \BibitemOpen
  \bibfield  {author} {\bibinfo {author} {\bibfnamefont {B.-A.}\ \bibnamefont
  {Li}}, \bibinfo {author} {\bibfnamefont {P.~G.}\ \bibnamefont {Krastev}},
  \bibinfo {author} {\bibfnamefont {D.-H.}\ \bibnamefont {Wen}}, \ and\
  \bibinfo {author} {\bibfnamefont {N.-B.}\ \bibnamefont {Zhang}},\ }\href
  {\doibase 10.1140/epja/i2019-12780-8} {\bibfield  {journal} {\bibinfo
  {journal} {Eur. Phys. J. A}\ }\textbf {\bibinfo {volume} {55}},\ \bibinfo
  {pages} {117} (\bibinfo {year} {2019})},\ \Eprint
  {http://arxiv.org/abs/1905.13175} {arXiv:1905.13175 [nucl-th]} \BibitemShut
  {NoStop}%
\bibitem [{\citenamefont {Wen}\ and\ \citenamefont {Holt}(2021)}]{Wen:2020nqs}%
  \BibitemOpen
  \bibfield  {author} {\bibinfo {author} {\bibfnamefont {P.}~\bibnamefont
  {Wen}}\ and\ \bibinfo {author} {\bibfnamefont {J.~W.}\ \bibnamefont {Holt}},\
  }\href {\doibase 10.1103/PhysRevC.103.064002} {\bibfield  {journal} {\bibinfo
   {journal} {Phys. Rev. C}\ }\textbf {\bibinfo {volume} {103}},\ \bibinfo
  {pages} {064002} (\bibinfo {year} {2021})},\ \Eprint
  {http://arxiv.org/abs/2012.02163} {arXiv:2012.02163 [nucl-th]} \BibitemShut
  {NoStop}%
\bibitem [{\citenamefont {Drischler}\ \emph {et~al.}(2020)\citenamefont
  {Drischler}, \citenamefont {Furnstahl}, \citenamefont {Melendez},\ and\
  \citenamefont {Phillips}}]{Drischler:2020hwi}%
  \BibitemOpen
  \bibfield  {author} {\bibinfo {author} {\bibfnamefont {C.}~\bibnamefont
  {Drischler}}, \bibinfo {author} {\bibfnamefont {R.~J.}\ \bibnamefont
  {Furnstahl}}, \bibinfo {author} {\bibfnamefont {J.~A.}\ \bibnamefont
  {Melendez}}, \ and\ \bibinfo {author} {\bibfnamefont {D.~R.}\ \bibnamefont
  {Phillips}},\ }\href {\doibase 10.1103/PhysRevLett.125.202702} {\bibfield
  {journal} {\bibinfo  {journal} {Phys. Rev. Lett.}\ }\textbf {\bibinfo
  {volume} {125}},\ \bibinfo {pages} {202702} (\bibinfo {year} {2020})},\
  \Eprint {http://arxiv.org/abs/2004.07232} {arXiv:2004.07232 [nucl-th]}
  \BibitemShut {NoStop}%
\bibitem [{\citenamefont {Somasundaram}\ \emph {et~al.}(2021)\citenamefont
  {Somasundaram}, \citenamefont {Drischler}, \citenamefont {Tews},\ and\
  \citenamefont {Margueron}}]{Somasundaram:2020chb}%
  \BibitemOpen
  \bibfield  {author} {\bibinfo {author} {\bibfnamefont {R.}~\bibnamefont
  {Somasundaram}}, \bibinfo {author} {\bibfnamefont {C.}~\bibnamefont
  {Drischler}}, \bibinfo {author} {\bibfnamefont {I.}~\bibnamefont {Tews}}, \
  and\ \bibinfo {author} {\bibfnamefont {J.}~\bibnamefont {Margueron}},\ }\href
  {\doibase 10.1103/PhysRevC.103.045803} {\bibfield  {journal} {\bibinfo
  {journal} {Phys. Rev. C}\ }\textbf {\bibinfo {volume} {103}},\ \bibinfo
  {pages} {045803} (\bibinfo {year} {2021})},\ \Eprint
  {http://arxiv.org/abs/2009.04737} {arXiv:2009.04737 [nucl-th]} \BibitemShut
  {NoStop}%
\bibitem [{\citenamefont {Imam}\ \emph {et~al.}(2022)\citenamefont {Imam},
  \citenamefont {Patra}, \citenamefont {Mondal}, \citenamefont {Malik},\ and\
  \citenamefont {Agrawal}}]{Imam:2021dbe}%
  \BibitemOpen
  \bibfield  {author} {\bibinfo {author} {\bibfnamefont {S.~M.~A.}\
  \bibnamefont {Imam}}, \bibinfo {author} {\bibfnamefont {N.~K.}\ \bibnamefont
  {Patra}}, \bibinfo {author} {\bibfnamefont {C.}~\bibnamefont {Mondal}},
  \bibinfo {author} {\bibfnamefont {T.}~\bibnamefont {Malik}}, \ and\ \bibinfo
  {author} {\bibfnamefont {B.~K.}\ \bibnamefont {Agrawal}},\ }\href {\doibase
  10.1103/PhysRevC.105.015806} {\bibfield  {journal} {\bibinfo  {journal}
  {Phys. Rev. C}\ }\textbf {\bibinfo {volume} {105}},\ \bibinfo {pages}
  {015806} (\bibinfo {year} {2022})},\ \Eprint
  {http://arxiv.org/abs/2110.15776} {arXiv:2110.15776 [nucl-th]} \BibitemShut
  {NoStop}%
\bibitem [{\citenamefont {Drischler}\ \emph {et~al.}(2021)\citenamefont
  {Drischler}, \citenamefont {Holt},\ and\ \citenamefont
  {Wellenhofer}}]{Drischler:2021kxf}%
  \BibitemOpen
  \bibfield  {author} {\bibinfo {author} {\bibfnamefont {C.}~\bibnamefont
  {Drischler}}, \bibinfo {author} {\bibfnamefont {J.~W.}\ \bibnamefont {Holt}},
  \ and\ \bibinfo {author} {\bibfnamefont {C.}~\bibnamefont {Wellenhofer}},\
  }\href {\doibase 10.1146/annurev-nucl-102419-041903} {\bibfield  {journal}
  {\bibinfo  {journal} {Ann. Rev. Nucl. Part. Sci.}\ }\textbf {\bibinfo
  {volume} {71}},\ \bibinfo {pages} {403} (\bibinfo {year} {2021})},\ \Eprint
  {http://arxiv.org/abs/2101.01709} {arXiv:2101.01709 [nucl-th]} \BibitemShut
  {NoStop}%
\bibitem [{\citenamefont {Sun}\ \emph {et~al.}(2024)\citenamefont {Sun},
  \citenamefont {Bhattiprolu},\ and\ \citenamefont {Lattimer}}]{Sun:2023xkg}%
  \BibitemOpen
  \bibfield  {author} {\bibinfo {author} {\bibfnamefont {B.}~\bibnamefont
  {Sun}}, \bibinfo {author} {\bibfnamefont {S.}~\bibnamefont {Bhattiprolu}}, \
  and\ \bibinfo {author} {\bibfnamefont {J.~M.}\ \bibnamefont {Lattimer}},\
  }\href {\doibase 10.1103/PhysRevC.109.055801} {\bibfield  {journal} {\bibinfo
   {journal} {Phys. Rev. C}\ }\textbf {\bibinfo {volume} {109}},\ \bibinfo
  {pages} {055801} (\bibinfo {year} {2024})},\ \Eprint
  {http://arxiv.org/abs/2311.00843} {arXiv:2311.00843 [nucl-th]} \BibitemShut
  {NoStop}%
\bibitem [{\citenamefont {Providencia}\ and\ \citenamefont
  {Rabhi}(2013)}]{Providencia:2012rx}%
  \BibitemOpen
  \bibfield  {author} {\bibinfo {author} {\bibfnamefont {C.}~\bibnamefont
  {Providencia}}\ and\ \bibinfo {author} {\bibfnamefont {A.}~\bibnamefont
  {Rabhi}},\ }\href {\doibase 10.1103/PhysRevC.87.055801} {\bibfield  {journal}
  {\bibinfo  {journal} {Phys. Rev. C}\ }\textbf {\bibinfo {volume} {87}},\
  \bibinfo {pages} {055801} (\bibinfo {year} {2013})},\ \Eprint
  {http://arxiv.org/abs/1212.5911} {arXiv:1212.5911 [nucl-th]} \BibitemShut
  {NoStop}%
\bibitem [{\citenamefont {Provid\^encia}\ \emph {et~al.}(2014)\citenamefont
  {Provid\^encia}, \citenamefont {Avancini}, \citenamefont {Cavagnoli},
  \citenamefont {Chiacchiera}, \citenamefont {Ducoin}, \citenamefont {Grill},
  \citenamefont {Margueron}, \citenamefont {Menezes}, \citenamefont {Rabhi},\
  and\ \citenamefont {Vida\~na}}]{Providencia:2013dsa}%
  \BibitemOpen
  \bibfield  {author} {\bibinfo {author} {\bibfnamefont {C.}~\bibnamefont
  {Provid\^encia}}, \bibinfo {author} {\bibfnamefont {S.~S.}\ \bibnamefont
  {Avancini}}, \bibinfo {author} {\bibfnamefont {R.}~\bibnamefont {Cavagnoli}},
  \bibinfo {author} {\bibfnamefont {S.}~\bibnamefont {Chiacchiera}}, \bibinfo
  {author} {\bibfnamefont {C.}~\bibnamefont {Ducoin}}, \bibinfo {author}
  {\bibfnamefont {F.}~\bibnamefont {Grill}}, \bibinfo {author} {\bibfnamefont
  {J.}~\bibnamefont {Margueron}}, \bibinfo {author} {\bibfnamefont {D.~P.}\
  \bibnamefont {Menezes}}, \bibinfo {author} {\bibfnamefont {A.}~\bibnamefont
  {Rabhi}}, \ and\ \bibinfo {author} {\bibfnamefont {I.}~\bibnamefont
  {Vida\~na}},\ }\href {\doibase 10.1140/epja/i2014-14044-7} {\bibfield
  {journal} {\bibinfo  {journal} {Eur. Phys. J. A}\ }\textbf {\bibinfo {volume}
  {50}},\ \bibinfo {pages} {44} (\bibinfo {year} {2014})},\ \Eprint
  {http://arxiv.org/abs/1307.1436} {arXiv:1307.1436 [nucl-th]} \BibitemShut
  {NoStop}%
\bibitem [{\citenamefont {Bednarek}\ \emph {et~al.}(2019)\citenamefont
  {Bednarek}, \citenamefont {Sladkowski},\ and\ \citenamefont
  {Syska}}]{Bednarek:2019xytNI}%
  \BibitemOpen
  \bibfield  {author} {\bibinfo {author} {\bibfnamefont {I.}~\bibnamefont
  {Bednarek}}, \bibinfo {author} {\bibfnamefont {J.}~\bibnamefont
  {Sladkowski}}, \ and\ \bibinfo {author} {\bibfnamefont {J.}~\bibnamefont
  {Syska}},\ }\href {\doibase 10.5506/APhysPolB.50.1849} {\bibfield  {journal}
  {\bibinfo  {journal} {Acta Phys. Polon. B}\ }\textbf {\bibinfo {volume}
  {50}},\ \bibinfo {pages} {1849} (\bibinfo {year} {2019})}\BibitemShut
  {NoStop}%
\bibitem [{\citenamefont {Mekjian}(2007{\natexlab{a}})}]{Mekjian:2007zz}%
  \BibitemOpen
  \bibfield  {author} {\bibinfo {author} {\bibfnamefont {A.~Z.}\ \bibnamefont
  {Mekjian}},\ }\href {\doibase 10.1016/j.physletb.2007.05.061} {\bibfield
  {journal} {\bibinfo  {journal} {Phys. Lett. B}\ }\textbf {\bibinfo {volume}
  {651}},\ \bibinfo {pages} {33} (\bibinfo {year} {2007}{\natexlab{a}})},\
  \Eprint {http://arxiv.org/abs/0712.1168} {arXiv:0712.1168 [nucl-th]}
  \BibitemShut {NoStop}%
\bibitem [{\citenamefont {Mekjian}(2007{\natexlab{b}})}]{Mekjian:2007mz}%
  \BibitemOpen
  \bibfield  {author} {\bibinfo {author} {\bibfnamefont {A.}~\bibnamefont
  {Mekjian}},\ }\href {\doibase 10.1209/0295-5075/80/22002} {\bibfield
  {journal} {\bibinfo  {journal} {Eur. Phys. Lett.}\ }\textbf {\bibinfo
  {volume} {80}},\ \bibinfo {pages} {22002} (\bibinfo {year}
  {2007}{\natexlab{b}})},\ \Eprint {http://arxiv.org/abs/0712.1778}
  {arXiv:0712.1778 [nucl-th]} \BibitemShut {NoStop}%
\bibitem [{\citenamefont {Nakano}\ and\ \citenamefont
  {Nishijima}(1953)}]{Nakano:1953zz}%
  \BibitemOpen
  \bibfield  {author} {\bibinfo {author} {\bibfnamefont {T.}~\bibnamefont
  {Nakano}}\ and\ \bibinfo {author} {\bibfnamefont {K.}~\bibnamefont
  {Nishijima}},\ }\href {\doibase 10.1143/PTP.10.581} {\bibfield  {journal}
  {\bibinfo  {journal} {Prog. Theor. Phys.}\ }\textbf {\bibinfo {volume}
  {10}},\ \bibinfo {pages} {581} (\bibinfo {year} {1953})}\BibitemShut
  {NoStop}%
\bibitem [{\citenamefont {Gell-Mann}(1956)}]{Gell-Mann:1956iqa}%
  \BibitemOpen
  \bibfield  {author} {\bibinfo {author} {\bibfnamefont {M.}~\bibnamefont
  {Gell-Mann}},\ }\href {\doibase 10.1007/BF02748000} {\bibfield  {journal}
  {\bibinfo  {journal} {Nuovo Cim.}\ }\textbf {\bibinfo {volume} {4}},\
  \bibinfo {pages} {848} (\bibinfo {year} {1956})}\BibitemShut {NoStop}%
\bibitem [{\citenamefont {Dexheimer}\ and\ \citenamefont
  {Schramm}(2008)}]{Dexheimer:2008ax}%
  \BibitemOpen
  \bibfield  {author} {\bibinfo {author} {\bibfnamefont {V.}~\bibnamefont
  {Dexheimer}}\ and\ \bibinfo {author} {\bibfnamefont {S.}~\bibnamefont
  {Schramm}},\ }\href {\doibase 10.1086/589735} {\bibfield  {journal} {\bibinfo
   {journal} {Astrophys. J.}\ }\textbf {\bibinfo {volume} {683}},\ \bibinfo
  {pages} {943} (\bibinfo {year} {2008})},\ \Eprint
  {http://arxiv.org/abs/0802.1999} {arXiv:0802.1999 [astro-ph]} \BibitemShut
  {NoStop}%
\bibitem [{\citenamefont {Dexheimer}\ and\ \citenamefont
  {Schramm}(2010)}]{Dexheimer:2009hi}%
  \BibitemOpen
  \bibfield  {author} {\bibinfo {author} {\bibfnamefont {V.~A.}\ \bibnamefont
  {Dexheimer}}\ and\ \bibinfo {author} {\bibfnamefont {S.}~\bibnamefont
  {Schramm}},\ }\href {\doibase 10.1103/PhysRevC.81.045201} {\bibfield
  {journal} {\bibinfo  {journal} {Phys. Rev. C}\ }\textbf {\bibinfo {volume}
  {81}},\ \bibinfo {pages} {045201} (\bibinfo {year} {2010})},\ \Eprint
  {http://arxiv.org/abs/0901.1748} {arXiv:0901.1748 [astro-ph.SR]} \BibitemShut
  {NoStop}%
\bibitem [{\citenamefont {Cruz-Camacho}\ \emph {et~al.}(2025)\citenamefont
  {Cruz-Camacho}, \citenamefont {Kumar}, \citenamefont {Reinke~Pelicer},
  \citenamefont {Peterson}, \citenamefont {Manning}, \citenamefont {Haas},
  \citenamefont {Dexheimer},\ and\ \citenamefont
  {Noronha-Hostler}}]{Cruz-Camacho:2024odu}%
  \BibitemOpen
  \bibfield  {author} {\bibinfo {author} {\bibfnamefont {N.}~\bibnamefont
  {Cruz-Camacho}}, \bibinfo {author} {\bibfnamefont {R.}~\bibnamefont {Kumar}},
  \bibinfo {author} {\bibfnamefont {M.}~\bibnamefont {Reinke~Pelicer}},
  \bibinfo {author} {\bibfnamefont {J.}~\bibnamefont {Peterson}}, \bibinfo
  {author} {\bibfnamefont {T.~A.}\ \bibnamefont {Manning}}, \bibinfo {author}
  {\bibfnamefont {R.}~\bibnamefont {Haas}}, \bibinfo {author} {\bibfnamefont
  {V.}~\bibnamefont {Dexheimer}}, \ and\ \bibinfo {author} {\bibfnamefont
  {J.}~\bibnamefont {Noronha-Hostler}} (\bibinfo {collaboration} {MUSES}),\
  }\href {\doibase 10.1103/PhysRevD.111.094030} {\bibfield  {journal} {\bibinfo
   {journal} {Phys. Rev. D}\ }\textbf {\bibinfo {volume} {111}},\ \bibinfo
  {pages} {094030} (\bibinfo {year} {2025})},\ \Eprint
  {http://arxiv.org/abs/2409.06837} {arXiv:2409.06837 [nucl-th]} \BibitemShut
  {NoStop}%
\bibitem [{\citenamefont {Aryal}\ \emph {et~al.}(2020)\citenamefont {Aryal},
  \citenamefont {Constantinou}, \citenamefont {Farias},\ and\ \citenamefont
  {Dexheimer}}]{Aryal:2020ocm}%
  \BibitemOpen
  \bibfield  {author} {\bibinfo {author} {\bibfnamefont {K.}~\bibnamefont
  {Aryal}}, \bibinfo {author} {\bibfnamefont {C.}~\bibnamefont {Constantinou}},
  \bibinfo {author} {\bibfnamefont {R.~L.~S.}\ \bibnamefont {Farias}}, \ and\
  \bibinfo {author} {\bibfnamefont {V.}~\bibnamefont {Dexheimer}},\ }\href
  {\doibase 10.1103/PhysRevD.102.076016} {\bibfield  {journal} {\bibinfo
  {journal} {Phys. Rev. D}\ }\textbf {\bibinfo {volume} {102}},\ \bibinfo
  {pages} {076016} (\bibinfo {year} {2020})},\ \Eprint
  {http://arxiv.org/abs/2004.03039} {arXiv:2004.03039 [nucl-th]} \BibitemShut
  {NoStop}%
\bibitem [{sup()}]{supplemental}%
  \BibitemOpen
  \href@noop {} {}\bibinfo {note} {See Supplemental Material at
  https://arxiv.org/abs/2504.18764 for the chemical potential in different
  bases, the derivation of symmetry energy expansion for both strange matter in
  weak equilibrium and isospin symmetric matter, and their fitting
  procedures.}\BibitemShut {Stop}%
\bibitem [{\citenamefont {Aryal}\ \emph {et~al.}(2021)\citenamefont {Aryal},
  \citenamefont {Constantinou}, \citenamefont {Farias},\ and\ \citenamefont
  {Dexheimer}}]{Aryal:2021ojz}%
  \BibitemOpen
  \bibfield  {author} {\bibinfo {author} {\bibfnamefont {K.}~\bibnamefont
  {Aryal}}, \bibinfo {author} {\bibfnamefont {C.}~\bibnamefont {Constantinou}},
  \bibinfo {author} {\bibfnamefont {R.~L.~S.}\ \bibnamefont {Farias}}, \ and\
  \bibinfo {author} {\bibfnamefont {V.}~\bibnamefont {Dexheimer}},\ }\href
  {\doibase 10.3390/universe7110454} {\bibfield  {journal} {\bibinfo  {journal}
  {Universe}\ }\textbf {\bibinfo {volume} {7}},\ \bibinfo {pages} {454}
  (\bibinfo {year} {2021})},\ \Eprint {http://arxiv.org/abs/2109.14787}
  {arXiv:2109.14787 [nucl-th]} \BibitemShut {NoStop}%
\bibitem [{\citenamefont {Reinke~Pelicer}\ \emph {et~al.}(2025)\citenamefont
  {Reinke~Pelicer} \emph {et~al.}}]{ReinkePelicer:2025vuh}%
  \BibitemOpen
  \bibfield  {author} {\bibinfo {author} {\bibfnamefont {M.}~\bibnamefont
  {Reinke~Pelicer}} \emph {et~al.},\ }\href@noop {} {\  (\bibinfo {year}
  {2025})},\ \Eprint {http://arxiv.org/abs/2502.07902} {arXiv:2502.07902
  [nucl-th]} \BibitemShut {NoStop}%
\bibitem [{\citenamefont {Bombaci}\ and\ \citenamefont
  {Lombardo}(1991)}]{Bombaci:1991zz}%
  \BibitemOpen
  \bibfield  {author} {\bibinfo {author} {\bibfnamefont {I.}~\bibnamefont
  {Bombaci}}\ and\ \bibinfo {author} {\bibfnamefont {U.}~\bibnamefont
  {Lombardo}},\ }\href {\doibase 10.1103/PhysRevC.44.1892} {\bibfield
  {journal} {\bibinfo  {journal} {Phys. Rev. C}\ }\textbf {\bibinfo {volume}
  {44}},\ \bibinfo {pages} {1892} (\bibinfo {year} {1991})}\BibitemShut
  {NoStop}%
\bibitem [{\citenamefont {Roca-Maza}\ \emph {et~al.}(2018)\citenamefont
  {Roca-Maza}, \citenamefont {Col\`o},\ and\ \citenamefont
  {Sagawa}}]{Roca-Maza:2018bpv}%
  \BibitemOpen
  \bibfield  {author} {\bibinfo {author} {\bibfnamefont {X.}~\bibnamefont
  {Roca-Maza}}, \bibinfo {author} {\bibfnamefont {G.}~\bibnamefont {Col\`o}}, \
  and\ \bibinfo {author} {\bibfnamefont {H.}~\bibnamefont {Sagawa}},\ }\href
  {\doibase 10.1103/PhysRevLett.120.202501} {\bibfield  {journal} {\bibinfo
  {journal} {Phys. Rev. Lett.}\ }\textbf {\bibinfo {volume} {120}},\ \bibinfo
  {pages} {202501} (\bibinfo {year} {2018})},\ \Eprint
  {http://arxiv.org/abs/1803.09120} {arXiv:1803.09120 [nucl-th]} \BibitemShut
  {NoStop}%
\bibitem [{\citenamefont {Danhoni}\ \emph {et~al.}(2025)\citenamefont
  {Danhoni}, \citenamefont {Yang}, \citenamefont {Hippert},\ and\ \citenamefont
  {Noronha-Hostler}}]{Danhoni:2025qpn}%
  \BibitemOpen
  \bibfield  {author} {\bibinfo {author} {\bibfnamefont {I.}~\bibnamefont
  {Danhoni}}, \bibinfo {author} {\bibfnamefont {Y.}~\bibnamefont {Yang}},
  \bibinfo {author} {\bibfnamefont {M.}~\bibnamefont {Hippert}}, \ and\
  \bibinfo {author} {\bibfnamefont {J.}~\bibnamefont {Noronha-Hostler}},\
  }\href@noop {} {\  (\bibinfo {year} {2025})},\ \Eprint
  {http://arxiv.org/abs/2510.23984} {arXiv:2510.23984 [nucl-th]} \BibitemShut
  {NoStop}%
\bibitem [{\citenamefont {Parotto}\ \emph {et~al.}(2020)\citenamefont
  {Parotto}, \citenamefont {Bluhm}, \citenamefont {Mroczek}, \citenamefont
  {Nahrgang}, \citenamefont {Noronha-Hostler}, \citenamefont {Rajagopal},
  \citenamefont {Ratti}, \citenamefont {Sch{\"a}fer},\ and\ \citenamefont
  {Stephanov}}]{Parotto:2018pwx}%
  \BibitemOpen
  \bibfield  {author} {\bibinfo {author} {\bibfnamefont {P.}~\bibnamefont
  {Parotto}}, \bibinfo {author} {\bibfnamefont {M.}~\bibnamefont {Bluhm}},
  \bibinfo {author} {\bibfnamefont {D.}~\bibnamefont {Mroczek}}, \bibinfo
  {author} {\bibfnamefont {M.}~\bibnamefont {Nahrgang}}, \bibinfo {author}
  {\bibfnamefont {J.}~\bibnamefont {Noronha-Hostler}}, \bibinfo {author}
  {\bibfnamefont {K.}~\bibnamefont {Rajagopal}}, \bibinfo {author}
  {\bibfnamefont {C.}~\bibnamefont {Ratti}}, \bibinfo {author} {\bibfnamefont
  {T.}~\bibnamefont {Sch{\"a}fer}}, \ and\ \bibinfo {author} {\bibfnamefont
  {M.}~\bibnamefont {Stephanov}},\ }\href {\doibase
  10.1103/PhysRevC.101.034901} {\bibfield  {journal} {\bibinfo  {journal}
  {Phys. Rev. C}\ }\textbf {\bibinfo {volume} {101}},\ \bibinfo {pages}
  {034901} (\bibinfo {year} {2020})},\ \Eprint
  {http://arxiv.org/abs/1805.05249} {arXiv:1805.05249 [hep-ph]} \BibitemShut
  {NoStop}%
\bibitem [{\citenamefont {Kahangirwe}\ \emph {et~al.}(2024)\citenamefont
  {Kahangirwe}, \citenamefont {Bass}, \citenamefont {Bratkovskaya},
  \citenamefont {Jahan}, \citenamefont {Moreau}, \citenamefont {Parotto},
  \citenamefont {Price}, \citenamefont {Ratti}, \citenamefont {Soloveva},\ and\
  \citenamefont {Stephanov}}]{Kahangirwe:2024cny}%
  \BibitemOpen
  \bibfield  {author} {\bibinfo {author} {\bibfnamefont {M.}~\bibnamefont
  {Kahangirwe}}, \bibinfo {author} {\bibfnamefont {S.~A.}\ \bibnamefont
  {Bass}}, \bibinfo {author} {\bibfnamefont {E.}~\bibnamefont {Bratkovskaya}},
  \bibinfo {author} {\bibfnamefont {J.}~\bibnamefont {Jahan}}, \bibinfo
  {author} {\bibfnamefont {P.}~\bibnamefont {Moreau}}, \bibinfo {author}
  {\bibfnamefont {P.}~\bibnamefont {Parotto}}, \bibinfo {author} {\bibfnamefont
  {D.}~\bibnamefont {Price}}, \bibinfo {author} {\bibfnamefont
  {C.}~\bibnamefont {Ratti}}, \bibinfo {author} {\bibfnamefont
  {O.}~\bibnamefont {Soloveva}}, \ and\ \bibinfo {author} {\bibfnamefont
  {M.}~\bibnamefont {Stephanov}},\ }\href {\doibase
  10.1103/PhysRevD.109.094046} {\bibfield  {journal} {\bibinfo  {journal}
  {Phys. Rev. D}\ }\textbf {\bibinfo {volume} {109}},\ \bibinfo {pages}
  {094046} (\bibinfo {year} {2024})},\ \Eprint
  {http://arxiv.org/abs/2402.08636} {arXiv:2402.08636 [nucl-th]} \BibitemShut
  {NoStop}%
\bibitem [{NA6(2022)}]{NA62KLEVER:2022nea}%
  \BibitemOpen
  in\ \href@noop {} {\emph {\bibinfo {booktitle} {{Snowmass 2021}}}}\ (\bibinfo
  {year} {2022})\ \Eprint {http://arxiv.org/abs/2204.13394} {arXiv:2204.13394
  [hep-ex]} \BibitemShut {NoStop}%
\bibitem [{\citenamefont {Haidenbauer}\ and\ \citenamefont
  {Meissner}(2010)}]{Haidenbauer:2009qn}%
  \BibitemOpen
  \bibfield  {author} {\bibinfo {author} {\bibfnamefont {J.}~\bibnamefont
  {Haidenbauer}}\ and\ \bibinfo {author} {\bibfnamefont {U.~G.}\ \bibnamefont
  {Meissner}},\ }\href {\doibase 10.1016/j.physletb.2010.01.031} {\bibfield
  {journal} {\bibinfo  {journal} {Phys. Lett. B}\ }\textbf {\bibinfo {volume}
  {684}},\ \bibinfo {pages} {275} (\bibinfo {year} {2010})},\ \Eprint
  {http://arxiv.org/abs/0907.1395} {arXiv:0907.1395 [nucl-th]} \BibitemShut
  {NoStop}%
\bibitem [{\citenamefont {Li}\ \emph {et~al.}(2018)\citenamefont {Li},
  \citenamefont {Hyodo},\ and\ \citenamefont {Geng}}]{Li:2018tbt}%
  \BibitemOpen
  \bibfield  {author} {\bibinfo {author} {\bibfnamefont {K.-W.}\ \bibnamefont
  {Li}}, \bibinfo {author} {\bibfnamefont {T.}~\bibnamefont {Hyodo}}, \ and\
  \bibinfo {author} {\bibfnamefont {L.-S.}\ \bibnamefont {Geng}},\ }\href
  {\doibase 10.1103/PhysRevC.98.065203} {\bibfield  {journal} {\bibinfo
  {journal} {Phys. Rev. C}\ }\textbf {\bibinfo {volume} {98}},\ \bibinfo
  {pages} {065203} (\bibinfo {year} {2018})},\ \Eprint
  {http://arxiv.org/abs/1809.03199} {arXiv:1809.03199 [nucl-th]} \BibitemShut
  {NoStop}%
\bibitem [{\citenamefont {Song}\ \emph {et~al.}(2018)\citenamefont {Song},
  \citenamefont {Li},\ and\ \citenamefont {Geng}}]{Song:2018qqm}%
  \BibitemOpen
  \bibfield  {author} {\bibinfo {author} {\bibfnamefont {J.}~\bibnamefont
  {Song}}, \bibinfo {author} {\bibfnamefont {K.-W.}\ \bibnamefont {Li}}, \ and\
  \bibinfo {author} {\bibfnamefont {L.-S.}\ \bibnamefont {Geng}},\ }\href
  {\doibase 10.1103/PhysRevC.97.065201} {\bibfield  {journal} {\bibinfo
  {journal} {Phys. Rev. C}\ }\textbf {\bibinfo {volume} {97}},\ \bibinfo
  {pages} {065201} (\bibinfo {year} {2018})},\ \Eprint
  {http://arxiv.org/abs/1802.04433} {arXiv:1802.04433 [nucl-th]} \BibitemShut
  {NoStop}%
\bibitem [{\citenamefont {Haidenbauer}\ \emph {et~al.}(2016)\citenamefont
  {Haidenbauer}, \citenamefont {Mei\ss{}ner},\ and\ \citenamefont
  {Petschauer}}]{Haidenbauer:2015zqb}%
  \BibitemOpen
  \bibfield  {author} {\bibinfo {author} {\bibfnamefont {J.}~\bibnamefont
  {Haidenbauer}}, \bibinfo {author} {\bibfnamefont {U.-G.}\ \bibnamefont
  {Mei\ss{}ner}}, \ and\ \bibinfo {author} {\bibfnamefont {S.}~\bibnamefont
  {Petschauer}},\ }\href {\doibase 10.1016/j.nuclphysa.2016.01.006} {\bibfield
  {journal} {\bibinfo  {journal} {Nucl. Phys. A}\ }\textbf {\bibinfo {volume}
  {954}},\ \bibinfo {pages} {273} (\bibinfo {year} {2016})},\ \Eprint
  {http://arxiv.org/abs/1511.05859} {arXiv:1511.05859 [nucl-th]} \BibitemShut
  {NoStop}%
\bibitem [{\citenamefont {Kumar}\ \emph {et~al.}(2024)\citenamefont {Kumar}
  \emph {et~al.}}]{MUSES:2023hyz}%
  \BibitemOpen
  \bibfield  {author} {\bibinfo {author} {\bibfnamefont {R.}~\bibnamefont
  {Kumar}} \emph {et~al.} (\bibinfo {collaboration} {MUSES}),\ }\href {\doibase
  10.1007/s41114-024-00049-6} {\bibfield  {journal} {\bibinfo  {journal}
  {Living Rev. Rel.}\ }\textbf {\bibinfo {volume} {27}},\ \bibinfo {pages} {3}
  (\bibinfo {year} {2024})},\ \Eprint {http://arxiv.org/abs/2303.17021}
  {arXiv:2303.17021 [nucl-th]} \BibitemShut {NoStop}%
\bibitem [{\citenamefont {Yang}\ \emph {et~al.}(2025)\citenamefont {Yang},
  \citenamefont {Hippert}, \citenamefont {Speranza},\ and\ \citenamefont
  {Noronha}}]{Yang:2025yoo}%
  \BibitemOpen
  \bibfield  {author} {\bibinfo {author} {\bibfnamefont {Y.}~\bibnamefont
  {Yang}}, \bibinfo {author} {\bibfnamefont {M.}~\bibnamefont {Hippert}},
  \bibinfo {author} {\bibfnamefont {E.}~\bibnamefont {Speranza}}, \ and\
  \bibinfo {author} {\bibfnamefont {J.}~\bibnamefont {Noronha}},\ }\href
  {\doibase 10.1103/9nck-dm66} {\bibfield  {journal} {\bibinfo  {journal}
  {Phys. Rev. Lett.}\ }\textbf {\bibinfo {volume} {135}},\ \bibinfo {pages}
  {222702} (\bibinfo {year} {2025})},\ \Eprint
  {http://arxiv.org/abs/2504.07805} {arXiv:2504.07805 [nucl-th]} \BibitemShut
  {NoStop}%
\bibitem [{\citenamefont {Harris}(2025)}]{Harris:2025ncu}%
  \BibitemOpen
  \bibfield  {author} {\bibinfo {author} {\bibfnamefont {S.~P.}\ \bibnamefont
  {Harris}},\ }\href {\doibase 10.1103/r28j-pvpm} {\bibfield  {journal}
  {\bibinfo  {journal} {Phys. Rev. C}\ }\textbf {\bibinfo {volume} {112}},\
  \bibinfo {pages} {035806} (\bibinfo {year} {2025})},\ \Eprint
  {http://arxiv.org/abs/2505.12133} {arXiv:2505.12133 [nucl-th]} \BibitemShut
  {NoStop}%
\end{thebibliography}%

\clearpage
\onecolumngrid
\section{Chemical potentials in different bases}

When dealing with strange baryon matter, one may use baryon number $B$, electric charge $Q$ and strangeness $\mathcal{S}$ as a basis of independent charges. Or one may use the $(B,I_z,S)$ basis, in terms of the isospin $I_z$ along the proton--neutron direction. 
However, when changing between the $(B,Q,S)$ and the $(B,I_z,S)$ bases, the
 chemical potentials transform in a non-trivial manner, and one must be careful about how they are defined \cite{Aryal:2020ocm}. 

In terms of $B$, $Q$ and $S$,
the first law of thermodynamics is given by
\begin{equation}
      dE = 
    \mu_B\, dB + \mu_Q\, dQ + \mu_S\, d\mathcal{S} + \ldots,
\end{equation}
where $E$ in the internal energy and we define the chemical potentials 
\begin{align}
    & \mu_B\equiv \left(\frac{\partial E}{\partial B}\right)_{Q,\mathcal{S}}, &
    & \mu_Q\equiv\left(\frac{\partial E}{\partial Q}\right)_{B,\mathcal{S}}, &
    & \mu_S\equiv \left(\frac{\partial E}{\partial \mathcal{S}}\right)_{B,Q},  &
\end{align}
We can also write the first law in terms of $B$, $I_z$ and $S$, by using
the the Gell-Mann--Nishijima formula $Q = I_z + (B+\mathcal{S})/2$:
\begin{align}
\begin{split}
    dE ={}& 
    \mu_B\, dB + \mu_Q \left( dI_z + \frac{1}{2}dB + \frac{1}{2}d\mathcal{S}\right) + \mu_S \,d\mathcal{S}+ \ldots
    \\={}&
    \mu_B^{(I S)}\, dB + \mu_I^{(B S)}\, dI_z + \mu_S^{(B I)}\, d\mathcal{S}+ \ldots,
\label{eq:appfirstlaw}
\end{split}
\end{align}
where we now define 
\begin{align}
  &
    \mu_B^{(I S)} \equiv \left(\frac{\partial E}{\partial B}\right)_{I_z,\mathcal{S}}
    ,
  &
  &
    \mu_I^{(B S)} \equiv \left(\frac{\partial E}{\partial I_z}\right)_{B,\mathcal{S}}
    ,
  &  
  &
    \mu_S^{(B I)} \equiv \left(\frac{\partial E}{\partial S}\right)_{B,I_z}
    .
  &  
\end{align}
The chemical potentials in the $(B,I_z,\mathcal{S})$ basis can be related to $\mu_B$, $\mu_S$ and $\mu_Q$ by inspecting Eq.~\eqref{eq:appfirstlaw}:
\begin{align}
  &
    \mu_B^{(I S)} 
    = \mu_B + \frac{1}{2}\mu_Q
    ,
  &
  &
    \mu_I^{(B S)}
    = \mu_Q
    ,
  &  
  &
    \mu_S^{(B I)} 
    = \mu_S + \frac{1}{2}\mu_Q
  &  
\end{align}

Notice that, even though $B$ and $S$ are the same regardless of whether $Q$ or $I_z$ is used, their respective chemical potentials  depend on which other quantity is fixed while they are varied. 
This could easily lead to confusion if one is not careful about notation. 
This point was recently discussed in Ref.~\cite{Aryal:2020ocm}.

\subsection{Matrix notation}

The Gell-Mann--Nishijima formula can be written in terms of a matrix $G$:
\begin{align}
&
\left(\begin{array}{c}
  B  \\
  Q  \\
  \mathcal{S}
\end{array}\right)
=G
\left(\begin{array}{c}
  B  \\
  I_z  \\
  \mathcal{S}
\end{array}\right),
&
&
\left(\begin{array}{c}
  \mu_B^{(I_z S)}  \\
 \mu_I^{(B S)}  \\
  \mu_S^{(B I_z)}
\end{array}\right)
=G^T
\left(\begin{array}{c}
  \mu_B\\
  \mu_Q\\
  \mu_S
\end{array}\right),
&
& \text{with} &
&
G \equiv 
\left(\begin{array}{ccc}
  1 & 0 & 0  \\
  \frac{1}{2} & 1 & \frac{1}{2}   \\
  0 & 0 & 1
\end{array}\right),
&
\end{align}
where chemical potentials transform so as to preserve the energy differential
\begin{equation}
    dE =
    \left(
\begin{array}{ccc}
    \mu_B   & \mu_Q & \mu_S
\end{array}
    \right)
      \left(
\begin{array}{c}
    dB \\ dQ \\ d\mathcal{S}
\end{array}
    \right)  
    +\ldots
    =
    \left(
\begin{array}{ccc}
    \mu_B^{(I_z S)}   
    & \mu_I^{(B S)} 
    & \mu_S^{(B I_z)}
\end{array}
    \right)
      \left(
\begin{array}{c}
    dB \\ dI_z \\ d\mathcal{S}
\end{array}
    \right) +\ldots.
\end{equation}

\subsection{Isospin reversal symmetry}

Isospin reflection symmetry is an element of $U(1)_B\times SU(3)_f$ which consists in flipping the isospin component $I_z$ in the proton-neutron direction while holding $B$ and $\mathcal{S}$ fixed:
\begin{align}
&
R_z:
&
&
\left(\begin{array}{c}
  B  \\
  I_z  \\
  \mathcal {S}
\end{array}\right)
\to 
\left(\begin{array}{c}
  B  \\
  I_z'  \\
  \mathcal{S}
\end{array}\right)
=
R_z^{(B I S)}
\left(\begin{array}{c}
  B  \\
  I_z  \\
  \mathcal{S}
\end{array}\right),
&
&
R_z^{(B I S)}\equiv
\left(\begin{array}{ccc}
  1 & 0 & 0  \\
 0 & -1 & 0   \\
  0 & 0 & 1
\end{array}\right)
.
&
\end{align}
We can express this transformation in the $(B,Q,S)$ basis by employing the matrix $G$:
\begin{align}
&
R_z:
&
&
\left(\begin{array}{c}
  B  \\
  Q  \\
  \mathcal{S}
\end{array}\right)
\to 
\left(\begin{array}{c}
  B'  \\
  Q'  \\
  \mathcal{S}'
\end{array}\right)
=
R_z^{(BQS)}
\left(\begin{array}{c}
  B  \\
  Q  \\
  \mathcal{S}
\end{array}\right),
&
&
R_z^{(BQS)} \equiv 
G
\cdot R_z^{(BIS)} \cdot 
G^{-1}
=
\left(\begin{array}{ccc}
  1 & 0 & 0  \\
 1 & -1 & 1   \\
  0 & 0 & 1
\end{array}\right),
&
\label{eq:appisorefBQS}
\end{align}
where we find that, because  $[R_z^{(BIS)},G]\neq 0$, isospin reflection symmetry is not even diagonal in the $(B,Q,S)$ basis, and is  very different from taking $Q\to-Q$. 
Moreover, the chemical potentials in the $(B,Q,S)$ basis transform with $(R_z^{(BQS)})^T$, so that isospin reflections affect both $\mu_B\to \mu_B + \mu_Q$ and $\mu_S\to \mu_S+\mu_Q$.

\section{Case 1: Symmetry-energy expansion with strangeness for $\mu_S=1/2\mu_Q$ (isospin-reflection--symmetric)}

\begin{figure}
    \centering
    \includegraphics[width=0.5\linewidth]{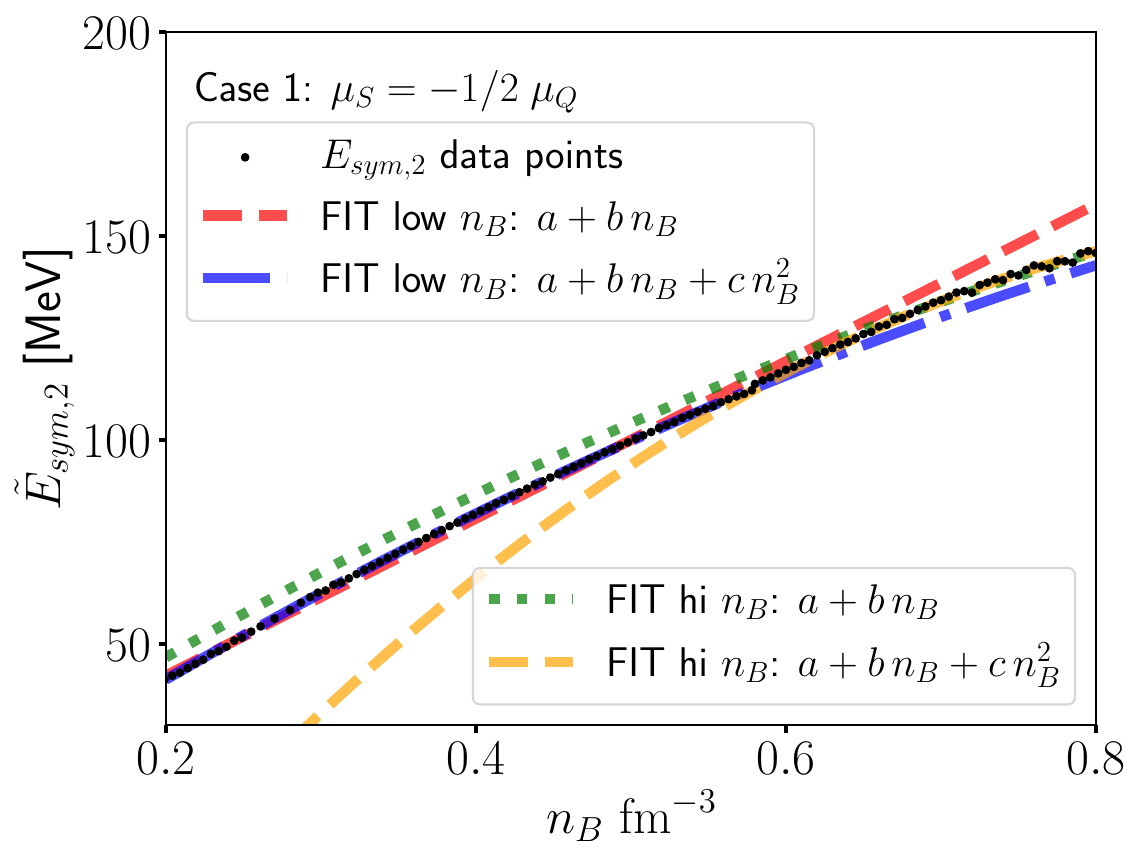}
    \caption{Symmetry energy vs $n_B$ in the isospin reflection symmetry case where $Y_S\neq0$. Various polynomial fits are used either at $n_{sat}$ (i.e. low $n_B$) or after the point strangeness appears (i.e. high $n_B$). We find that a fit at $n_{sat}$ up to $n_B^2$ reproduces CMF++ the best.}
    \label{fig:Esym2_sym}
\end{figure}

In the main text, we already saw in Fig.\ \ref{fig:EA_muI} that the isospin-relection--symmetric case of $\mu_S=1/2\mu_Q$ could be well described using only $\delta_I^2$ terms. 
Then, it remains to be seen what the $n_B$ dependence of the symmetry energy $\tilde{E}_{sym,2}(n_B)$ is in the presence of strangeness. 
In Fig.\ \ref{fig:Esym2_sym} we plot $\tilde{E}_{sym,2}(n_B)$ from CMF++ compared to different types of polynomial fits of different orders where we either expand around $n_{sat}$ or a point at high $n_B$ where $Y_S\neq 0$. 
We find that a polynomial up to $n_B^2$, expanded around $n_{sat}$ is the clear winner. 
Thus, we may write down our full symmetry-energy expansion for the isospin-relection--symmetric case with strangeness:
\begin{eqnarray}
    \frac{\tilde{E}_{ANM}}{A}(n_B,\delta_I)\Big|_{\mu_S=\frac{\mu_Q}{2}}=\frac{\tilde{E}_{SNM}}{A}(n_B)+\left[E_{sym2,sat}+L_{sym2,sat}\left(\frac{n_B-n_{sat}}{n_{sat}}\right)+K_{sym2,sat}\left(\frac{n_B-n_{sat}}{n_{sat}}\right)^2\right]\,\delta_I^2 , 
\end{eqnarray}
where all expansion coefficients are absorbed in the terms $\left\{E_{sym2,sat},L_{sym2,sat},K_{sym2,sat}\right\}$. 

\section{Case 2: Symmetry-energy expansion with strangeness for $\mu_S=0$ (isospin reflection symmetry breaking)}

Next we consider the case of weak equilibrium where we know there is isospin reflection symmetry breaking due to the skewness of $Y_S(\delta_I)$. 
As we discussed already in the main text, weak equilibrium leads to a finite $\delta^3$ dependence of the symmetry-energy expansion such that the general form is 
\begin{eqnarray}\label{eqn:weak_symE}
\frac{\tilde{E}_{ANM}}{A}(n_B,\delta)\Big|_{\mu_S=0}&=&\frac{\tilde{E}_{SNM}}{A}(n_B)+\tilde{E}_{sym,2}(n_B)\,\delta^2  +\tilde{E}_{sym,3}(n_B)\,\delta^3
\end{eqnarray}
where the $\tilde{E}_{sym,3}(n_B)$ is only non-zero after the onset of strangeness. 
In Fig.\ \ref{fig:fit_order_CMF} we plot both $\tilde{E}_{sym,2}(n_B)$ and $\tilde{E}_{sym,3}(n_B)$ that we have extracted from CMF++ in the case of weak equilibrium. 
We find that not only does a skewness term appear i.e. $\tilde{E}_{sym,3}(n_B)\neq 0$ after the onset of strangeness but also  $\tilde{E}_{sym,2}(n_B)$ experiences a jump as well. 
Thus, it is clear that in the case of weak equilibrium that two expansion points are required: one around $n_{sat}$ as well as a second point that we call $n_B^{sc}$ that is after the onset of strangeness. 
Additionally, we compare different polynomial fits to the CMF++ using the C2 coupling results and find that in all cases a fit up to $n_B^2$ most accurately represents the data.

Returning to Eq.\ (\ref{eqn:deltaEXPAN}), we include only the relevant and nonzero terms:
\
\begin{eqnarray}
S(n_B,\delta_I)&=&\tilde{E}_{sym,2}(n_B)\delta^2 +\tilde{E}_{sym,3}(n_B)\delta^3\nonumber\\
&=&\left[ \tilde{E}_{sym,2}^{sat}(n_B)+\tilde{E}_{sym,2}^{sc}(n_B)\right]\delta^2 +\tilde{E}_{sym,3}(n_B)\delta^3, \nonumber\\
\end{eqnarray}
where these terms dependence on $n_B$ is:
\begin{subequations}
\begin{eqnarray}
    \tilde{E}_{sym,2}^{sat}(n_B)&=&E_{\textrm{sym}2,sat}+L_{sym2,sat}\frac{n_B-n_B^{sat}}{n_B^{sat}}+\frac{K_{sym2,sat}}{2}\left(\frac{n_B-n_B^{sat}}{n_B^{sat}}\right)^2, \\
    \tilde{E}_{sym,2}^{sc}(n_B)&=&\left[E_{\textrm{sym}2,sc}+L_{sym2,sc}\frac{n_B-n_B^{sc}}{n_B^{sc}}+\frac{K_{sym2,sc}}{2}\left(\frac{n_B-n_B^{sc}}{n_B^{sc}}\right)^2\right]\theta(n_B-n_B^{sc}), \\
\tilde{E}_{sym,3}(n_B)&=&\left[E_{\textrm{sym}3,sc}+L_{sym3,sc}\frac{n_B-n_B^{sc}}{n_B^{sc}}+\frac{K_{sym3,sc}}{2}\left(\frac{n_B-n_B^{sc}}{n_B^{sc}}\right)^2\right]\theta(n_B-n_B^{sc}).
\end{eqnarray}
\end{subequations}
Our new strangeness symmetry-energy expansion reduces to 3 coefficients found at $n_{sat}$ $\left\{E_{\textrm{sym}2,sat},L_{sym2,sat},K_{sym2,sat}\right\}$ and 6 coefficients $\left\{E_{\textrm{sym}2,sc},L_{sym2,sc},K_{sym2,sc},E_{\textrm{sym}3,sc},L_{sym3,sc},K_{sym3,sc}\right\}$
found at $n_{sc}$, leading to a total of 9 unknown coefficients.

At $n_B$ above the strangeness threshold we obtain a significant contribution from:
\begin{equation}
    \tilde{E}_{sym,2}^{sc}(n_B)\equiv\left[\tilde{E}_{sym,2}(n_B)-\tilde{E}_{sym,2}^{sat}(n_B)\right]\theta(n_B-n_B^{sc}), 
\end{equation}
where 
\begin{equation}
    \theta(n_B-n_B^{sc})=\left\{ \begin{matrix}
        0 \quad n_B<n_B^{sc}\\
         1 \quad n_B\geq n_B^{sc}
    \end{matrix}\right. , 
\end{equation}
and the Heaviside function ensures that the second expansion only occurs after strangeness is non-zero. Here, we set $n_B^{sc}$ as the density where strangeness becomes finite at the ground state. Due to the phase transition of various orders in CMF++, for a small window after the strangeness turns on, the fit struggles to reproduce the actual data (see Fig. \ref{fig:fit_order_CMF}).

In principle, it would be nice to have $n_B^{sc}$ correspond to the exact point when strangeness becomes finite (and this may be possible in certain models). In practice we found it best to choose $n_B^{sc}$  to be slightly above that point because the consequence of strangeness switching on can occur at a phase transition of various order in CMF++, which produces significant noise in the expansion if fitted  (e.g. see Fig. \ref{fig:fit_order_CMF} specifically for $\tilde{E}_{sym,3}(n_B)$). 

We now turn to our cubic term where we find here an expansion up to $\delta_I^3$ is sufficient for a reasonable description of the CMF++ EOS. 
Then, the cubic term can be defined as:
\begin{eqnarray}
   \tilde{E}_{sym,3}(n_B)\equiv \frac{S\left(n_B,\delta_I\right)-\tilde{E}_{sym,2}(n_B)\delta_I^2}{\delta_I^3}\theta(n_B-n_B^{sc}),
\end{eqnarray}
where our expansion point is once again $(n_B-n_B^{sc})/n_B^{sc}$ because the skewness in the symmetry energy only appears in the regime where strangeness is nonzero.

\subsection{Fitting Procedure to extra symmetry-energy coefficients}

At this point, we have not defined the values of these coefficients for our new symmetry-energy expansion.
We point out here, that in the original  symmetry-energy expansion, one defines two limits: SNM where $\delta_Q=0$ and PNM where $\delta_Q=1$. 
The original symmetry-energy expansion, then all coefficients were defined in terms of the two limits of SNM vs PNM. For instance, $E_{\textrm{sym}2,sat}$ would then be:
\begin{eqnarray}
E_{\textrm{sym}2,sat}^{org}&=&\left[\frac{\tilde{E}_{ANM}(\delta_Q)}{A}-\frac{\tilde{E}_{SNM}}{A}\right]\Big/\delta_Q^2 \nonumber \\
&=&\frac{\tilde{E}_{PNM}}{A}-\frac{\tilde{E}_{SNM}}{A}, 
\end{eqnarray}
which makes sense in models that PNM is a reasonable limit to their calculations and $\delta_Q$ is bounded by 1.
\emph{However}, once one redefines $\delta_I$ then it is no longer bounded by 1 due to the existence of multistrange baryons. 
Furthermore, for certain $n_B$ reaching $\delta_I\rightarrow 1$ may be challenging because strangeness can have a nontrivial interplay with $Y_Q$ (especially since s quarks carrying $Q_s=-1/3$ electric charge whereas strange hadrons may be positive, negative, or neutral).  
Additionally, values of $\delta_I= 1$ do not imply PNM. 
In fact, it implies only that $Y_Q=Y_S/2$, which could be achieved with a variety of combinations of particles from the baryon octet and decuplet.

Thus, we introduce a new method to extract the coefficients using polynomial fits to $S(n_B,\delta)$.  
At a fixed $n_B=const$, we fit a polynomial to the energy per nucleon $E/A$ of the form
\begin{equation}
    f_{E/A}(\delta)=a_0+a_1\delta+a_2\delta^2+a_3\delta^3+a_4\delta^3, 
\end{equation}
where we find that $a_1=0$ as long as we use $\delta_I$ instead of $\delta_Q$. 
Additionally, for $n_B<n_B^{sc}$, we also find that $a_3=0$. 
We have checked the influence of higher-order contributions and found that $a_4$ is negligible. 
The results of such a fit are shown in Fig.\ \ref{fig:skewness} in the main text. 
Thus, we are left with two functions:
\begin{eqnarray}
    \text{If } n_B<n_B^{sc} &\quad& f_{E/A}(\delta)=a_0+a_2\delta^2,\\
    \text{else }  &\quad& f_{E/A}(\delta)=b_0+b_2\delta^2+b_3\delta^3.
\end{eqnarray}
Then, this process is repeated across all $n_B$ such that we construction functions of $a_i(n_B)$ and $b_i(n_B)$.
Then, $\tilde{E}_{sym,2}^{sat}(n_B)=a_2(n_B)$ where the symmetry-energy coefficients can be extracted by fitting a polynomial of the form:
\begin{equation}
    \text{If } n_B<n_B^{sc} \quad f_{sym,2,sat}(n_B)= E_{\textrm{sym}2,sat}+L_{sym2,sat}\left(\frac{n_B-n_{sat}}{n_{sat}}\right)+\frac{K_{sym2,sat}}{2}\left(\frac{n_B-n_{sat}}{n_{sat}}\right)^2+\dots, 
\end{equation}
to $a_2(n_B)$. 
For CMF++ we find that the $n_{sat}$ expansion only requires at most $\left\{E_{\textrm{sym}2,sat},L_{sym2,sat},K_{sym2,sat}\right\}$ to provide a reasonable fit (such that we drop higher-order terms like $J_{sym2,sat}$). 
The comparison of our fitting procedure to the true CMF++ EOS is shown in Fig.\ \ref{fig:fit_order_CMF} (left) where the comparison relevant here is at low $n_B$.

Above $n_B>n_B^{sc}$ more care needs to be taken because the expansion is complex. 
Similarly to what we did before, we can simply set the skewness term as $\tilde{E}_{sym,3}^{sat}(n_B)=b_3(n_B)$, since it has no contribution from the regime around $n_{sat}$. 
Then we fit a polynomial of the form
\begin{equation}
    \text{If } n_B>n_B^{sc} \quad f_{sym,3,sc}(n_B)=E_{\textrm{sym}3,sc}+L_{sym3,sc}\left(\frac{n_B-n_{sc}}{n_{sc}}\right)+K_{sym3,sc}\left(\frac{n_B-n_{sc}}{n_{sc}}\right)^2+\dots, 
\end{equation}
where the leading order term may be 0 in the case of a smooth transition into finite strangeness. 
However, here we keep all three coefficients for a reasonable description i.e. $\left\{E_{\textrm{sym}3,sc},L_{sym3,sc},K_{sym3,sc}\right\}$. 
The result of our fitting procedure is shown in Fig.\ \ref{fig:fit_order_CMF} (left) where our three terms provide a reasonable fit (but struggle to get the behavior precisely where strangeness becomes finite due to the phase transition). 

Finally, we still must obtain the coefficients for $\tilde{E}_{sym,2}^{sat}(n_B)$ but this is more challenging because our fitted $b_2(n_B)$ contains contributions both from $\tilde{E}_{sym,2}^{sat}(n_B)$ and $\tilde{E}_{sym,2}^{sc}(n_B)$.
Thus, we must set:
\begin{equation}
    \tilde{E}_{sym,2}^{sc}(n_B)=b_2(n_B)-\tilde{E}_{sym,2}^{sat}(n_B), 
\end{equation}
such that we are only extracting the difference of the 2nd order symmetry energy from the expansion around $n_{sat}$.
Then, after this subtraction, we can fit a polynomial to 
\begin{equation} \tilde{E}_{sym,2}^{sc}(n_B)=E_{\textrm{sym}2,sc}+L_{sym2,sc}\frac{n_B-n_B^{sc}}{n_B^{sc}}+\frac{K_{sym2,sc}}{2}\left(\frac{n_B-n_B^{sc}}{n_B^{sc}}\right)^2+\dots, 
\end{equation}
where once again there may be some scenarios where $E_{\textrm{sym}2,sc}\rightarrow 0$ if strangeness switches on very smoothly. However, we continue to keep all three coefficients in our fits $\left\{E_{\textrm{sym}2,sc},L_{sym2,sc},K_{sym2,sc}\right\}$. 
The result of this fit is shown in Fig.\ \ref{fig:fit_order_CMF} (left) where our fit is shown at high $n_B$.  
In Fig.\ \ref{fig:fit_order_CMF} (left) the two different fits from the low $n_B$ expansion around $n_{sat}$ are also shown extending upwards past their regime of validity. 
This extension past their regime of validity is to demonstrate the failure of the expansion around only  $n_{sat}$ in the presence of strangeness. 

In this work, we have compared our extraction of the symmetry-energy coefficients to more traditional methods that take derivatives of the EOS directly. We found that we are able to recover the original symmetry-energy coefficients of CMF++ around $n_{sat}$, but our method allows us to calculate even higher-order coefficients that were not possible before. 
Of course, with the polynomial expansion, there is a danger of incorporating infinitely many coefficients and over-fitting the function. 
Thus, we choose here to extract as few coefficients as reasonably possible to extract the symmetry-energy expansion. 
Finally, we note that further care must be taken when handling phase transitions into a finite strangeness.  
Here we were most interested in understanding if our new strangeness symmetry-energy expansion could generally reproduce the symmetry energy in CMF++. 
However, we suspect there are methods to better capture the more complicated regime as one switches from light to strangeness, which we plan to explore in future work.

\end{document}